%% file: B2G-24-001_temp.tex
\pdfoutput=1
\documentclass[11pt,twoside,a4paper,cmspaper,final,collab]{cms-tdr}
\def\svnVersion{5b34a14}\def\svnDate{2025/12/30}\def\cmsCernNoTag{CERN-EP-2025-160}\def\cmsCernDate{\today}\def\cmsMessage{Published in the Journal of High Energy Physics as \href{http://dx.doi.org/10.1007/JHEP12(2025)178}{\doi{10.1007/JHEP12(2025)178}.}}
\begin{document}\cmsNoteHeader{B2G-24-001}

\newcommand{\PY}{{\HepParticle{Y}{}{}}\xspace}
\newcommand{\mX}{\ensuremath{m_\PX}\xspace}
\newcommand{\mY}{\ensuremath{m_\PY}\xspace}
\newcommand{\mH}{\ensuremath{m_{\PH}}}
\newcommand{\mgg}{\ensuremath{m_{\PGg\PGg}}\xspace}
\newcommand{\TwoNLL}{\ensuremath{-2 \ln{L}}\xspace}
\newcommand{\gjet}{\ensuremath{\PGg + \text{jets}}\xspace}
\newcommand{\dipho}{\ensuremath{\PGg\PGg + \text{jets}}\xspace}
\newcommand{\vgamma}{\ensuremath{\PV + \PGg}\xspace}
\newcommand{\ttgamma}{\ensuremath{\ttbar + \PGg}\xspace}
\newcommand{\ttgammagamma}{\ensuremath{\ttbar + \PGg\PGg}\xspace}
\newcommand{\ttV}{\ensuremath{\ttbar\PV}\xspace}
\newcommand{\tZq}{\ensuremath{\PQt\PZ\PQq}\xspace}

\cmsNoteHeader{B2G-24-001}
\title{Search for a new scalar resonance decaying to a Higgs boson and another new scalar particle in the final state with two bottom quarks and two photons in proton-proton collisions at \texorpdfstring{$\sqrt{s}=13\TeV$}{sqrt(s) = 13 TeV}}

\date{\today}

\abstract{
A search is presented for a new scalar resonance, \PX, decaying to a standard model Higgs boson and another new scalar particle, \PY, in the final state where the Higgs boson decays to a $\PQb\PAQb$ pair, while the \PY particle decays to a pair of photons. The search is performed in the mass range 240--1000\GeV for the resonance \PX, and in the mass range 70--800\GeV for the particle \PY, using proton-proton collision data collected by the CMS experiment at $\sqrt{s}=13\TeV$, corresponding to an integrated luminosity of 132\fbinv. In general, the data are found to be compatible with the standard model expectation. Observed (expected) upper limits at 95\% confidence level on the product of the production cross section and the relevant branching fraction are extracted for the $\PX\to\PY\PH$ process, and are found to be within the range of 0.05--2.69\,(0.08--1.94)\unit{fb}, depending on \mX and \mY. The most significant deviation from the background-only hypothesis is observed for \PX and \PY masses of 300 and 77\GeV, respectively, with a local\,(global) significance of 3.33\,(0.65) standard deviations.
}

\hypersetup{
pdfauthor={CMS Collaboration},
pdftitle={Search for a new scalar resonance decaying to a Higgs boson and a new scalar with two bottom quarks and two photons in the final state in proton-proton collisions at sqrt(s) = 13 TeV},
pdfsubject={CMS},
pdfkeywords={CMS, Higgs, parametric neural network, photons, b quarks, BSM}}

\maketitle

\section{Introduction}
\label{sec:introduction}

Since the discovery of a Higgs boson by the ATLAS and CMS Collaborations at the CERN LHC in 2012~\cite{HiggsATLAS2012,HiggsCMS2012,Chatrchyan:2013lba}, an extensive campaign of experimental investigations has been launched to determine its properties.
The latest results from these studies have been published by the two collaborations in 2022~\cite{HiggsATLAS2022,HiggsCMS2022}.
The Higgs boson properties that have been measured with the current data sets are compatible with the standard model (SM) prediction.

Despite the progress in studying the characteristics of the Higgs boson, there are still some open questions related to its nature, \eg, the hierarchy problem---the question of why the Higgs mass is so much smaller than the Planck scale.
Also, the SM is not a complete theory as it has no explanations for neutrino masses and the origin of dark matter.
Such shortcomings could be addressed by theoretical models beyond the SM (BSM).

One of these BSM theories is supersymmetry (SUSY)~\cite{Ramond:1971gb,Golfand:1971iw,Neveu:1971rx,Volkov:1972jx,Wess:1973kz,Wess:1974tw,Fayet:1974pd,Nilles:1983ge}, which predicts the existence of partners for each SM particle having the same properties as their SM counterparts but differing in mass and spin.
By extending the SM in this way, SUSY offers a solution to the hierarchy problem, provides an explanation for the nature of dark matter, and modifies the running of the coupling constants of the fundamental forces so that they approximately converge to a common value at a very large energy scale, as predicted by grand unified theories.

In its simplest form, denoted as the minimal supersymmetric SM (MSSM)~\cite{MSSM}, SUSY requires the introduction of one additional Higgs $SU(2)$-doublet, leading to the prediction of five Higgs boson mass eigenstates, instead of only one as in the SM.
The MSSM Lagrangian contains a supersymmetric mass term for these Higgs doublets associated with a mass parameter, $\mu_\text{SUSY}$, that has to be at the scale of the SUSY breaking, $m_\text{SUSY}$, to explain the observed phenomenology.
However, there is no strong theoretical reason in the MSSM for $\mu_\text{SUSY}$ and $m_\text{SUSY}$ to be of the same order.
This ``necessary accident" of the MSSM is called the ``$\mu$-problem", and can be resolved in further extensions of the model.
More specifically, the next-to-MSSM (NMSSM)~\cite{NMSSM} introduces an additional Higgs singlet field, S, which provides an explanation for the origin and the scale of the $\mu_\text{SUSY}$ term.

Because of the existence of the extra singlet field, the NMSSM Higgs sector ends up having seven gauge eigenstates.
The gauge eigenstates combine to form seven mass eigenstates: three $CP$-even ($\PH_1$, $\PH_2$, $\PH_3$), two $CP$-odd (A$_1$, A$_2$), and two charged ($\PH^\pm$), where the subscripts indicate the mass ordering.
In the so-called alignment limit, the heavier neutral eigenstates only weakly couple to the 125\GeV Higgs boson (\PH) observed at the LHC, and, thus, their decay to $\PH\PH$ is suppressed.
In this scenario, decays of the form $\PX\to\PY\PH$, where X is one of the heaviest neutral eigenstates (H$_3$ or A$_2$) and Y is a lighter one, are preferred, and motivate the search presented in this paper.
A complete picture of the NMSSM theory is given in Ref.~\cite{NMSSM}.

Another category of BSM models with the potential to explain some of the SM shortcomings is the two-real-scalar-singlet model (TRSM).
More concretely, in the specific extension of TRSM detailed in Ref.~\cite{TRSM}, the asymmetric signature with one SM Higgs boson and one new scalar resonance (\PY) can occur.
The main difference with the NMSSM is that, in the TRSM, the additional scalars (\PX and \PY) have SM Higgs boson branching fractions ($\mathcal{B}$), which implies a small $\mathcal{B}(\PY\to\PGg\PGg)$, while in the NMSSM there is more freedom for this branching fraction to take a wide range of possible values.
Further details and benchmarks for the models discussed above are given in Ref.~\cite{benchmarks}.

We note that this search probes the $\PX\to\PY\PH$ final state in a model-independent way.
The X and Y particles are taken to be scalar, and we further assume that the narrow-width approximation is applicable.
The connection to the aforementioned models can be recovered by appropriately matching the \PX and \PY particles to scalars in the mass hierarchy of each model.

\begin{figure}[!hbtp]
    \centering
    \includegraphics[width=0.5\textwidth]{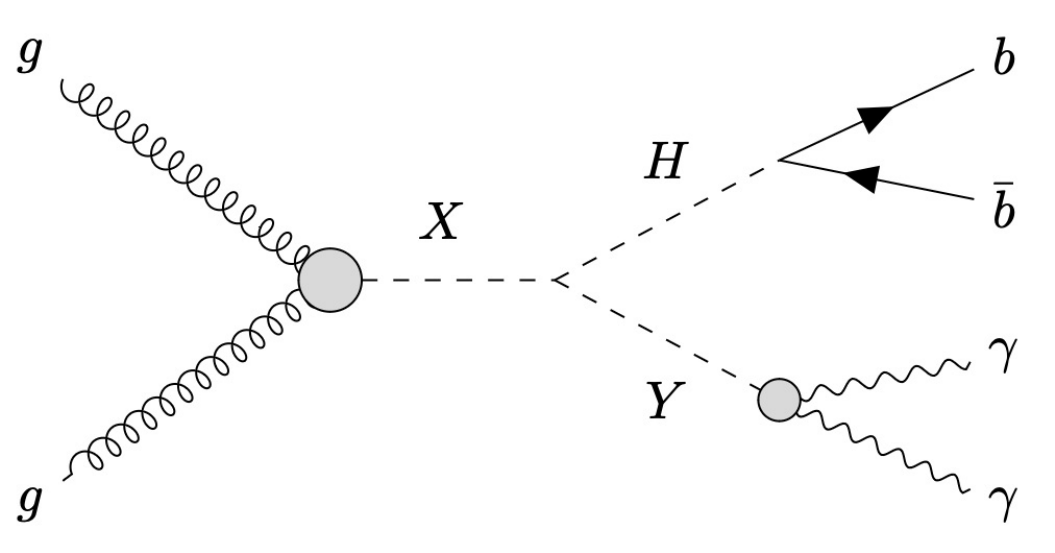}
\caption{Feynman diagram for the production of the BSM resonance \PX and its subsequent decay to two scalars, one SM Higgs boson and one BSM scalar \PY, with $\PH\to\PQb\PAQb$ and $\PY\to\PGg\PGg$.}
\label{fig:diagramXToYggHbb}
\end{figure}

Different combinations of $\PX\to\PY\PH$ final states have been studied by the CMS Collaboration:
$\PY\to\PQb\PAQb$ and $\PH\to\PQb\PAQb$\cite{XToYHTo4b_boosted};
$\PY\to\PQb\PAQb$ and $\PH\to\PGt\PGt$~\cite{XToYHTobbtautau};
$\PY\to\PGt\PGt$ and $\PH\to\PGg\PGg$, and $\PY\to\PGg\PGg$ and $\PH\to\PGt\PGt$~\cite{XToYHToggtautau};
as well as $\PY\to\PQb\PAQb$ and $\PH\to\PGg\PGg$~\cite{XToYbbHgg}.
The search described in this paper probes the so far unexplored $\PY\to\PGg\PGg$ and $\PH\to\PQb\PAQb$ final state (Fig.~\ref{fig:diagramXToYggHbb}) for masses of the \PX scalar between 240 and 1000\GeV and masses of the \PY scalar satisfying the condition $\mX>\mY+\mH$, with $\mY>70\GeV$.
The mass range is chosen based on several considerations, including detector sensitivity at low masses and the need for a different set of reconstruction techniques at higher mass, adjusted for boosted topologies.
Additionally, the condition $\mX>\mY+\mH$ ensures that the decay $\PX\to\PY H$ is kinematically allowed.
This final state is well motivated by the high $\mathcal{B}(\PH\to\PQb\PAQb)$ and the clean signature of the photon pair.

Apart from theoretical considerations, this search is also prompted by excesses observed in similar analyses searching for new resonances in signatures with Higgs bosons.
The search for $\PX\to\PY(\PQb\PAQb)\PH(\PGg\PGg)$ (same final state as this analysis but with inverted decays modes for \PY and \PH) has observed a local (global) excess of 3.8 (2.8) standard deviations for $\mX = 650\GeV$ and $\mY = 90\GeV$~\cite{XToYbbHgg}.
Under the assumption that the \PY can also decay to a pair of photons and that the branching fractions of $\PY\rightarrow\PGg\PGg$ and $\PY\rightarrow\PQb\PAQb$ are the same as those of the Higgs boson, as predicted in the TRSM, this excess can be probed by this search.
Moreover, a search for an additional, SM-like, low-mass Higgs boson in the $\PH\to\PGg\PGg$ channel found an excess of local (global) significance of 2.9 (1.3) standard deviations at 95.4\GeV~\cite{LowMassHgg}.

The paper is structured as follows:
Section~\ref{sec:CMSObjects} gives a description of the CMS detector and the event reconstruction.
The data sets and simulated samples used in the analysis are detailed in Section~\ref{sec:samples}.
Section~\ref{sec:analysisStrategy} describes general analysis strategy, the event selection and categorization, and Section~\ref{sec:modeling} outlines the modeling of the signal and the background.
The systematic uncertainties are discussed in Section~\ref{sec:systematics}.
Finally, results are presented in Section~\ref{sec:results}, and the analysis is summarized in Section~\ref{sec:summary}.
The results of the analysis are provided in tabulated form in the HEPData record~\cite{hepdata}.

\section{The CMS detector and event reconstruction}
\label{sec:CMSObjects}

The central feature of the CMS apparatus is a superconducting solenoid of 6\unit{m} internal diameter, providing a magnetic field of 3.8\unit{T}.
Within the solenoid volume are a silicon pixel and strip tracker, a lead tungstate crystal electromagnetic calorimeter (ECAL), and a brass and scintillator hadron calorimeter (HCAL), each composed of a barrel and two endcap sections.
Forward calorimeters extend the pseudorapidity coverage provided by the barrel and endcap detectors.
Muons are reconstructed in gas-ionization detectors embedded in the steel flux-return yoke outside the solenoid.
More detailed descriptions of the CMS detector, together with a definition of the coordinate system used and the relevant kinematic variables, can be found in Refs.~\cite{CMS:2008xjf,CMS:2023gfb}.

Events of interest are selected using a two-tiered trigger system. The first level (L1), composed of custom hardware processors, uses information from the calorimeters and muon detectors to select events at a rate of around 100\unit{kHz} within a fixed latency of about 4\mus~\cite{CMS:2020cmk}.
The second level, known as the high-level trigger (HLT), consists of a farm of processors running a version of the full event reconstruction software optimized for fast processing, and reduces the event rate to around 1\unit{kHz} before data storage~\cite{CMS:2016ngn,CMS:2024aqx}.

The primary vertex (PV) is taken to be the vertex corresponding to the hardest scattering in the event, evaluated using tracking information alone, as described in Section~9.4.1 of Ref.~\cite{CMS-TDR-15-02}.
As a consequence of the presence of jets from the $\PH\to\PQb\PAQb$ decay, this approach provides a similar or better \mgg resolution to that obtained in $\PH\to\PGg\PGg$ analyses using a customized vertex finding method.
A particle-flow (PF) algorithm~\cite{CMS:2017yfk} aims to reconstruct and identify each individual particle in the event, with an optimized combination of information from the various elements of the CMS detector in the following paragraphs.

The energy of photons is obtained from the ECAL measurement.
In the barrel section of the ECAL, an energy resolution of about 1\% is achieved for unconverted or late-converting photons in the tens of \GeVns{} energy range.
The energy resolution of the remaining barrel photons is about 1.3\% up to absolute pseudorapidity $\abs{\eta} = 1$, changing to about 2.5\% at $\abs{\eta} = 1.4$.
In the endcaps, the energy resolution is about 2.5\% for unconverted or late-converting photons, and between 3 and 4\% for the other ones~\cite{CMS:2015myp}.
The diphoton mass resolution, as measured in $\PH \to \PGg\PGg$ decays, is typically in the 1--2\% range, depending on the measurement of the photon energies in the ECAL and the topology of the photons in the event~\cite{CMS:2020xrn}.

The energy of electrons is determined from a combination of the electron momentum at the PV, as determined by the tracker, the energy of the corresponding ECAL cluster, and the energy sum of all bremsstrahlung photons spatially compatible with originating from the electron track.
The momentum resolution for electrons with transverse momentum $\pt \approx 45\GeV$ from $\PZ \to \Pep \Pem$ decays ranges from 2 to 5\%.
It is generally better in the barrel region than in the endcaps, and also depends on the bremsstrahlung energy emitted by the electron as it traverses the material in front of the ECAL~\cite{CMS:2020uim,CMS-DP-2020-021}.

The energy of muons is obtained from the curvature of the corresponding track.
Muons are measured in the pseudorapidity range $\abs{\eta} < 2.4$, with detection planes made using three technologies: drift tubes, cathode strip chambers, and resistive-plate chambers.
Matching muons to tracks measured in the silicon tracker results in a relative transverse momentum resolution, for muons with $20 < \pt < 100\GeV$, of 1.3--2.0\% in the barrel and better than 6\% in the endcaps.
Measurements made with cosmic ray muons show that, in the central region of the detector, the \pt resolution is better than 10\% for muons with \pt up to 1\TeV~\cite{Chatrchyan:2012xi}.

The energy of charged hadrons is determined from a combination of their momentum measured in the tracker and the matching ECAL and HCAL energy deposits, corrected for the response function of the calorimeters to hadronic showers.
The energy of neutral hadrons is obtained from the corresponding corrected ECAL and HCAL energies.

Jets are reconstructed offline using the anti-\kt algorithm~\cite{Cacciari:2008gp, Cacciari:2011ma} with a distance parameter of 0.4.
Jet momentum is determined as the vectorial sum of all PF object momenta in the jet, and is found from simulation to be, on average, within 5 to 10\% of the true momentum over the entire \pt spectrum and detector acceptance.
Pileup, referring to additional proton-proton (pp) interactions within the same or nearby bunch crossings (pileup), can contribute additional tracks and calorimetric energy depositions, thereby increasing the apparent jet momentum.
To mitigate this effect, tracks identified to be originating from pileup vertices are discarded and an offset correction is applied to correct for remaining contributions~\cite{CMS:2020ebo}.
Jet energy corrections are derived from simulation studies so that the average measured energy of jets becomes identical to that of particle-level jets.
In situ measurements of the momentum balance in dijet, $\text{photon} + \text{jet}$, $\PZ + \text{jet}$, and multijet events are used to determine any residual differences between the jet energy scale in data and in simulation, and appropriate corrections are made~\cite{CMS:2016lmd}.
Additional selection criteria are applied to each jet to remove jets potentially dominated by instrumental effects or reconstruction failures~\cite{CMS:2020ebo}.
The identification of \PQb quark jets uses the \textsc{DeepJet} algorithm~\cite{Bols:2020bkb,CMS-DP-2018-058}.
In this analysis, no explicit requirement is placed on the \textsc{DeepJet} score; instead the continuous distribution of the \textsc{DeepJet} score is used as an input to the multivariate discriminator used for the final event selection, as discussed in Sections~\ref{sec:selectionCategorization:preselection} and~\ref{sec:selectionCategorization:pNN}.

The PF isolation of an object is quantified by the \pt sum of other PF objects within a cone $\Delta R = \sqrt{\smash[b]{(\Delta\eta)^2+(\Delta\phi)^2}}$ around its direction, where $\phi$ is the azimuthal angle.
The cut-off value chosen for $\Delta R$ depends on the type of object.

\section{Data and simulated samples}
\label{sec:samples}

This search uses pp collision data recorded in 2016--2018 by the CMS experiment at the LHC.
The data are collected by L1 and HLT algorithms that require at least two photons satisfying $\pt>30\GeV$ for the leading- and $\pt>18\GeV$ for the subleading-$\pt$ photon, respectively, and a diphoton invariant mass $\mgg>55\GeV$.
Additional criteria are applied by the trigger algorithms on the photons based on the shower profiles in the calorimeter,
the ratio of the energy deposition in the HCAL to that in the ECAL, and the isolation of the photons from PF objects in the same event.
Details will be further discussed in Sec~\ref{sec:selectionCategorization:preselection}.
The integrated luminosity associated with the data set is $132\fbinv$.

In all simulated samples discussed below, parton showering, fragmentation with the standard \pt-ordered parton shower scheme, as well as the underlying event description, are modeled using \PYTHIA 8.240 with the CP5 tune~\cite{CMS:2015wcf, CMS:2019csb}.
The NNPDF 3.1 set~\cite{NNPDF31} is used to describe the parton distribution functions, and the \GEANTfour package~\cite{Agostinelli:2002hh} is used to model the detector response.

Only the gluon-gluon fusion process is considered for the production of \PX in the simulated signal samples. These samples are generated with \MGvATNLO v2.6.5 at leading order (LO) precision~\cite{MadGraph,MadGraphNLO}.
The new scalar bosons in the signal are simulated as $CP$-even, and the product of the signal production cross section and its branching fraction to the $\PY\to\PGg\PGg$ and $\PH\to\PQb\PAQb$ final state is normalized to $1\unit{fb}$.
As already mentioned, we assume that the narrow-width approximation holds, i.e., the particle decay width is small compared to its mass.
As a consequence, possible interference with any type of background process is ignored.

The SM production of single Higgs bosons decaying to a pair of photons
constitutes a resonant background for this search.
To model this background we utilize simulated samples for the production of Higgs bosons via gluon-gluon and vector boson fusion, as well as associated production of Higgs bosons with vector bosons or with a top quark-antiquark pair ($\ttbar\PH$).
In all of these samples the Higgs boson is forced to decay into two photons. These Monte Carlo samples are generated with \MGvATNLO v2.6.5 at NLO in quantum chromodynamics (QCD), with cross sections and decay branching fraction taken from Ref.~\cite{deFlorian:2016spz}.
Contributions from other single \PH production modes are negligible and, therefore, omitted.
Although less important, we also consider the contribution of gluon-gluon fusion Higgs pair production.
It is generated with the \POWHEG~2~\cite{Frixione:2007vw,Alioli:2009je,Melia:2011tj,Nason:2013ydw} program at next-to-leading order (NLO) precision in perturbative QCD~\cite{Bagnaschi:2011tu,Heinrich:2020ckp,Heinrich:2019bkc,Jones:2017giv,Heinrich:2017kxx}, including the full top quark mass dependence~\cite{Buchalla:2018yce}.

The rest of the resonant (Drell--Yan and dibosons) and all the nonresonant background processes are estimated from data.
However, simulated samples are used for the optimization of the analysis strategy.
The Drell--Yan (DY) simulated sample is generated using \POWHEG~2 at NLO in perturbative QCD, and the diboson $\PV\PV$, where $\PV=\PW,\PZ$, samples are generated using either \MGvATNLO v2.6.5 or \POWHEG~2 at NLO in QCD.
The irreducible nonresonant prompt diphoton process ($\dipho$) is modeled with \SHERPA v.2.2.1~\cite{Gleisberg:2008ta} at LO including the box processes, with up to three additional jets in the final state.
The reducible nonresonant background from \gjet events is modeled with \PYTHIA~8~\cite{Sjostrand:2014zea} at LO.
Less important nonresonant backgrounds include \vgamma, \ttgamma, \ttgammagamma, \ttV, and \tZq and are simulated at NLO in QCD using \MGvATNLO v2.6.5, except for \ttbar, which uses \MGvATNLO v2.6.1.

Pileup interactions are taken into account by adding simulated minimum bias interactions to all the generated event samples.
The pileup distribution is weighted to match that observed in the data.
The mean value of pileup multiplicity ranges from 23 to 32, depending on the year of data taking.

\section{Analysis strategy and event categorization}
\label{sec:analysisStrategy}

The analysis probes a wide range of signal hypotheses (mass points), and its strategy is built around the necessity to be sensitive for all of them.
The baseline selection, denoted as ``preselection" below, applies to all of the mass points.
It relies on a series of selection criteria on individual photons and photon pairs, and on loose requirements on individual jets and dijet pairs.
The preselected events are then given as input to a parametric neural network (PNN), described in detail in Section~\ref{sec:selectionCategorization:pNN}.
The PNN is able to produce a separate output score optimized for each mass point it has been trained on, as well as for intermediate mass points.
This dedicated output score is then used to split events in different signal regions (SRs), in order to create categories with high sensitivity for each mass point.
In these SRs, the signal manifests as a mass peak in the \mgg spectrum, at different values depending on the value of \mY for each mass point.
The signal and background models are fitted together and compared to the data.
In the absence of significant deviations from the SM  prediction, we set an upper limit on the product of the cross section and branching fraction at a wide range of mass points, exploiting the interpolating capabilities of the PNN.

\subsection{Event preselection}
\label{sec:selectionCategorization:preselection}

The analysis preselection largely follows the trigger selection.
The first step consists of selecting events with at least two photon candidates each with $\pt > 18\GeV$ and with calorimeter superclusters within the ECAL barrel or endcap acceptance.  To reduce backgrounds, we impose additional requirements on these photon candidates.
The ratio of the energy deposited in the HCAL to that in the ECAL must satisfy $H/E<0.08$, effectively suppressing the contribution from QCD processes.
To reduce backgrounds from DY and $\PV\PV$ processes, with electrons mimicking the photon signature, we apply a requirement that there cannot be a charged track reconstructed in the silicon pixel tracker that points to the photon cluster.
A photon candidate for which the associated supercluster is close to a track that is compatible with an electron is also rejected.
This further reduces the contamination from misidentified electron backgrounds.

Isolation requirements on photons are applied separately to the charged hadron ($I_\text{ch}$), photon ($I_\text{ph}$), and the track ($I_\text{tk}$) components of the photon PF isolation, calculated within a cone of $\Delta R=0.3$ centered around the photon.
Two other variables, $R_{9}$ and $\sigma_{i\eta i\eta}$, are commonly used when photons are present in the final state.
The variable $R_{9}$ is defined as the energy sum of the $3\times3$ crystals centered around the most energetic crystal, divided by the energy of the photon.
The variable $\sigma_{i\eta i\eta}$ is the energy-weighted spread in the $\eta$ direction of the $5\times5$ crystals centered around the most energetic crystal.
A series of requirements are applied on the above variables to identify the photon candidates in this analysis, matching the online selection.
Every photon is required to fulfill $I_\text{ch}<20\GeV$, or $I_\text{ch}/\pt<0.3$, or $R_{9}>0.8$.
Additional requirements are applied with different values based on the photon candidate $\abs{\eta}$ and $R_{9}$ variables, as summarized in Table~\ref{tab:photonID}.
These requirements provide approximately 60\% rejection for backgrounds with misidentified photons, with approximately 2\% loss of signal efficiency.

\begin{table}[htbp]
    \centering
    \topcaption{Additional photon requirements, as functions of $\abs{\eta}$ and $R_{9}$. The variable $\rho$ is the median of the transverse energy density per unit area in the event.}
    \label{tab:photonID}
    \begin{tabular}{ccccc}
        $\abs{\eta}$ & $R_{9}$ & $\sigma_{i\eta i\eta}$ & $I_\text{ph}~[\GeVns]$ & $I_\text{tk}~[\GeVns]$\\
        \hline
        \vphantom{} \vspace{-9pt} \\
        \multicolumn{1}{l}{\multirow{2}{*}{Barrel}} & [0.50,0.85] & $<$0.015 & ${<}4.0 + 0.17\rho$ & $<$6.0 \\
        & $>$0.85 & \NA & \NA & \NA \\
        \vphantom{} \vspace{-9pt} \\
        \multicolumn{1}{l}{\multirow{2}{*}{Endcap}} & [0.80,0.90] & $<$0.035 & ${<}4.0 + 0.13\rho$ & $<$6.0 \\
        & $>$0.90 & \NA & \NA & \NA \\
    \end{tabular}
\end{table}

Finally, a multivariate algorithm (``photon MVA ID") is used to further select photon candidates.
The photon MVA ID is a general-purpose identification variable designed to distinguish photons from jets. It is computed based on multiple photon characteristics that describe the properties of the photon electromagnetic shower and isolation~\cite{CMS:2020uim}.
We impose a photon MVA ID requirement (working point, WP) that has a 90\% efficiency for isolated photons relevant for this analysis.

All of the photon candidates in a given event are paired up in all possible combinations.
In each diphoton pair, the photon with the highest (lowest) \pt is referred to as the ``leading" (``subleading") photon.
The leading photon is required to have $\pt > 30\GeV$, and the invariant mass of the pair must be larger than 55\GeV.
After these selections, the diphoton pair with the highest sum \pt is selected as the diphoton pair to be used in the analysis.
In the signal, the selected diphoton pair is treated as the \PY resonance candidate.

The final state for the signal of interest does not include isolated leptons.
To suppress backgrounds with leptons in the final state, we reject events with at least one electron of $\pt>10\GeV$ or muon of $\pt>15\GeV$ satisfying the criteria described below.
Electrons and muons must be within detector acceptance, fulfill tight impact parameter criteria, and be separated from any photon candidate by a distance of $\Delta R > 0.2$.
Electrons must also satisfy a requirement on an MVA ID variable, which includes isolation information~\cite{CMS:2020uim}.
The WP of the electron MVA ID is chosen to correspond to the 90\% efficiency point.
Tight identification criteria are also imposed on muons~\cite{CMS:2018rym}.
Additionally, muon candidates must be reconstructed as ``global muons"~\cite{CMS:2018rym}, meaning that tracks associated with them must be found both in the tracker and the muon detectors.
Finally, the PF relative isolation for muons is required to be $<0.3$, which corresponds to $\sim\!98\%$ efficiency.

Events are required to have at least two jets with $\abs{\eta}<2.4$ and $\pt>25\GeV$, each separated from any photon candidate by a distance $\Delta R>0.4$.
No selection on the \textsc{DeepJet} \PQb tagging score is applied at the preselection stage, to allow the PNN to make a b-jet selection that relies on additional event variables and that can be optimized for each signal mass hypothesis.
Instead, jets are sorted according to this score, and the two leading ones based on this sorting are preselected.
A tighter jet selection, including additional kinematic requirements, for example on the dijet mass, is performed by the PNN classifier, as discussed in Section~\ref{sec:selectionCategorization:pNN}.
A correction is applied to the entirety of the \PQb tagging discriminant distribution in the simulation to match the one in data.
To reject events with severe jet-related mismeasurements due to the presence of cosmic ray muons, beam-gas interactions, or beam halo or calorimetric noise, events in both data and simulation are required to pass dedicated noise-rejecting filters~\cite{CMS:2019ctu}.

\subsection{Estimation of backgrounds with misidentified photons from data}
\label{sec:modeling:bkg:DDQCDGJets}
Typically, simulated samples have difficulties reproducing the behavior and the distributions of events with misidentified photons in data.
Moreover, given the large phase space covered by this analysis, the statistical power of samples simulating QCD and \gjet processes is not always sufficient to provide a representative background sample for the training of the MVA methods at the core of the signal versus background classification used in this analysis.

For these reasons, we implemented an estimation of processes with one or two misidentified photons from control samples in data~\cite{PhysRevLett.125.061801}.
This can be accomplished using data events in a misidentified-photon-enriched region, independent of the preselection described in Section~\ref{sec:selectionCategorization:preselection}, as a proxy for the QCD and \gjet events passing the preselection.
The proxy data set is constructed by selecting events that pass all the preselection requirements, but fail the minimum photon MVA ID score criterion.
The event properties used as input to the final MVA-based  selection are found to be uncorrelated with the minimum photon MVA ID variable.
This proxy data set can then  be used, with the small modification described below, in the training of the MVA discriminator described in Section~\ref{sec:selectionCategorization:pNN} (the list of training variables is summarized in Table~\ref{tab:training_features}).

In the proxy data set we replace the photon MVA ID score for the photon failing the photon MVA ID requirement with a new value. This new value is randomly chosen in the range from the lower photon MVA ID score threshold to the photon MVA ID score of the other photon in the event, according to the probability distribution function (PDF) that describes the photon MVA ID scores of misidentified photons, as determined from MC \gjet events.
Finally, to properly inject the data-driven description of the misidentified photons into the preselection, the sample normalization has to be adjusted.
This is done by applying an individual weight to each of the events in the proxy dataset to match the photon MVA ID distribution for events in the preselection.

This technique improves the overall agreement between simulated samples and data preselection yields, as well as the description of the individual kinematic distributions, resulting in a better background description for training of the MVA described in the following section.
The final estimation and modeling of the backgrounds from processes with misidentified photons is performed directly on the data from sideband regions, as discussed in Section~\ref{sec:modeling:bkg}, and hence it is independent of the method described in this section.

\subsection{Parametric neural network selection}
\label{sec:selectionCategorization:pNN}

The signal for this search involves two new resonances with unknown masses that are probed in the mass range of 240--1000\GeV.
This creates the necessity to derive optimal discriminators for the vastly different mass points being considered.
Given that it is not realistic to train a single NN for each mass point because of the large number of such points, a mass-point-aware NN that can generalize the selection optimally to all of the mass points is preferred.
This can be achieved with a PNN~\cite{Baldi_2016}, as described below.

An NN is a function, $f(\vec{x})$, that outputs a score reflecting the similarity of an event, described by the set of input features $\vec{x}$, to signal-like characteristics.
A set of such functions $\{f^1, f^2,...\}$, where $f^i$ is an NN trained on $(\mX^i,\mY^i)$, may be assembled to create $f(\vec{x}; \mX,\mY)$:
\begin{equation}
f\left(\vec{x}; \mX, \mY\right) =
\begin{cases}

    f^1(\vec{x}) & \quad \text{if }(\mX^1,\mY^1), \\
    f^2(\vec{x}) & \quad \text{if }(\mX^2,\mY^2), \\
    \ldots
  \end{cases}
\label{eqn:combined_fx}
\end{equation}
When training a PNN, the target function is $f(\vec{x}; \mX,\mY)$, i.e., a single function, which, given a value for the \mX-\mY pair, will provide an optimal discriminator for the chosen mass points.
When using this function, picking a value of \mX and \mY is equivalent to picking a particular NN to use.
This is achieved in practice by including \mX and \mY as additional input features.
The training of the PNN uses 50\% of the available events in each sample and is performed simultaneously on the 224 simulated signal samples corresponding to different \mX and \mY hypotheses.
All simulated background samples mentioned in Section~\ref{sec:samples} as well as the backgrounds with misidentified photons, estimated as described in Section~\ref{sec:modeling:bkg:DDQCDGJets}, are included in the training.
Each signal sample consists of $\mathcal{O}(10^5)$ events, and the number of total background events used in the training is $\mathcal{O}(10^7)$.
Different \mX and \mY values are randomly assigned to the background events to match the signal points included in training. This procedure ensures that \mX and \mY features have no discriminating power between signal and background.  Instead, the performance of the PNN per mass point is optimized through the correlation between mass values and the training variables.

In order not to bias the training towards a specific signal mass hypothesis, the signal samples are normalized such that the sum of weights is the same for all mass points. All of the signal mass hypotheses are collectively normalized so that their sum of weights is the same as the one for the background samples, to enhance the discrimination based on shape differences between signal and background.

The PNN used in this analysis is a feedforward network with 3 layers of 50 nodes.
The intermediate layers are activated by exponential linear unit (ELU) functions with a dropout probability of 0.05, while the last layer is activated by a sigmoid function.
The learning rate is set to 0.01.

The set of training variables given as input to the PNN consists of the most discriminating variables, identified from studies of separate boosted decision trees (BDTs) trained with a much larger set of variables on a representative subset of signal mass points.
The correlation of some of the kinematic variables, such as photon \pt, with the diphoton mass is minimized by dividing them by \mgg.
Any variable that is still correlated with the diphoton mass is removed from the set of training variables.
This prevents the introduction of peaky features to the otherwise smoothly falling \mgg distribution of the nonresonant backgrounds. These features would invalidate the background modeling procedure described in Section~\ref{sec:modeling:bkg}.
The final list of training variables is shown in Table~\ref{tab:training_features}.

\begin{table}[htbp]
    \centering
    \topcaption{The training variables included as input to the PNN used for the final selection of this search. The symbols $\PGg_1$, $\PGg_2$ denote the leading and subleading photons, while $j_1$ and $j_2$ denote the leading and subleading jets. }
    \begin{tabular}{ll}
        Category & Variables \\ \hline
        \vphantom{} \vspace{-9pt} \\
        Photon-related & $\pt(\PGg\PGg)/\mgg$, $\pt(\PGg_1)/\mgg$, $\pt(\PGg_2)/\mgg$, \\
        & $\Delta R(\PGg\PGg)$, $\Delta\eta(\PGg\PGg)$ \\
        \vphantom{} \vspace{-9pt} \\
        Jet-related &  $\pt(j_1)$, $\pt(jj)$,  $m(jj)$, $\Delta R(jj)$,  $m(j_1)$, \\
        & $\text{b tagging score}(j_1)$, $\text{b tagging score}(j_2)$, \\
        & $\Delta R(\PGg_1, j_1)$, $\Delta R(\PGg_1, j_2)$, $\Delta R(\PGg_2, j_1)$, $\Delta R(\PGg_2, j_2)$ \\
        \vphantom{} \vspace{-9pt} \\
        Photon-jet &  $\Delta R(\PGg_1, jj)$, $\Delta R(\PGg_2, jj)$, $\Delta R(\PGg\PGg, j_1)$, $\Delta R(\PGg\PGg, j_2)$, $\Delta R(\PGg\PGg, jj)$, \\
        combinations & $\Delta \eta(\PGg\PGg, jj), \Delta \eta(\PGg\PGg, j_1), \Delta \eta(\PGg_1, j_2), $ \\
        & $\Delta \phi(\PGg\PGg, jj), \Delta \phi(\PGg\PGg, j_1), \Delta \phi(\PGg\PGg, j_2)$
    \end{tabular}
    \label{tab:training_features}
\end{table}

Before defining the SRs of the search, the PNN output score is transformed for each mass point such that the score distribution of the background from the test data set is constant, \ie, ``flat".
The new score that results from this procedure is referred to as the ``transformed score" and is used for the event categorization below.
Placing a threshold on the transformed score is equivalent to placing a threshold on the background efficiency.

We note that simpler classification methods, such as BDTs, were used for a small set of representative mass points to understand the performance gains from using a PNN.
BDTs based on a few of the most discriminating variables, effectively emulating a cutoff-based analysis, showed a much worse performance for the majority of the mass points.
The BDTs using as input the full list of variables mentioned in Table~\ref{tab:training_features} improved the performance for all mass points, making it comparable to the one provided by the PNN up to around 5\%, but still slightly worse.
In summary, the PNN brings a small performance gain to the analysis with respect to a simpler BDT classification while also providing the unique feature of using a single training to optimally probe multiple mass points, including those never used in training.

\subsection{Event categorization}
\label{sec:selectionCategorization:categorization}

The PNN transformed score is used in the final event selection to define the SRs of the search.
The SR boundaries are chosen to optimize the expected upper limit on the $\PX\to\PY(\PGg\PGg)\PH(\PQb\PAQb)$ cross section based on a simplified signal and background modeling.
For illustration, the transformed score distributions in the SRs of the signal hypotheses of $(\mX,\mY) = (280,90)\GeV$ and $(\mX,\mY) = (600,70)\GeV$ are shown in Fig.~\ref{fig:pNNScore}.
The details of the optimization procedure are given below.
This method of optimization produced similar results to that of a method based on optimization of the significance.

\begin{figure}[!hbtp]
    \centering
    \includegraphics[width=0.49\textwidth]{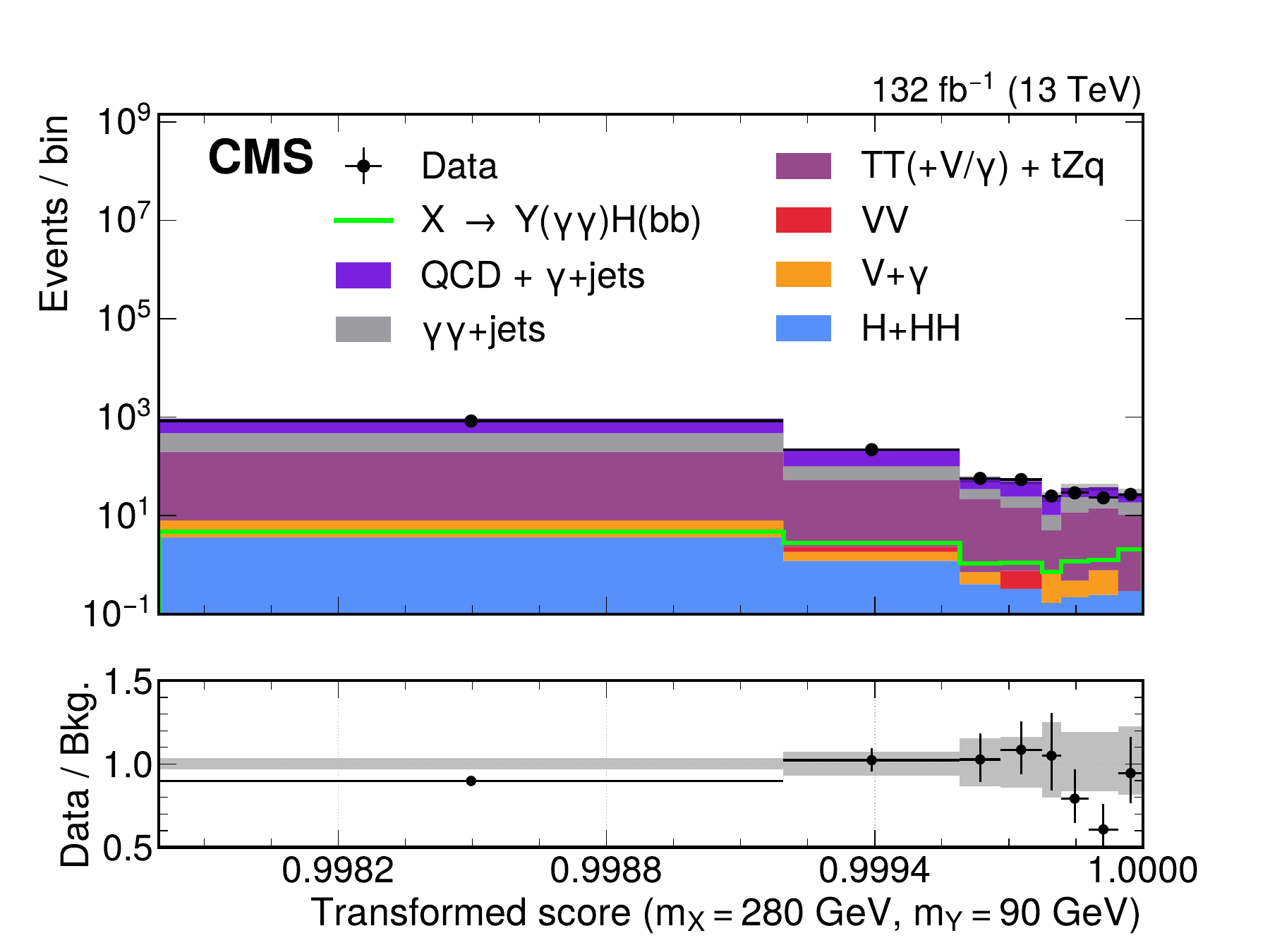}
    \includegraphics[width=0.49\textwidth]{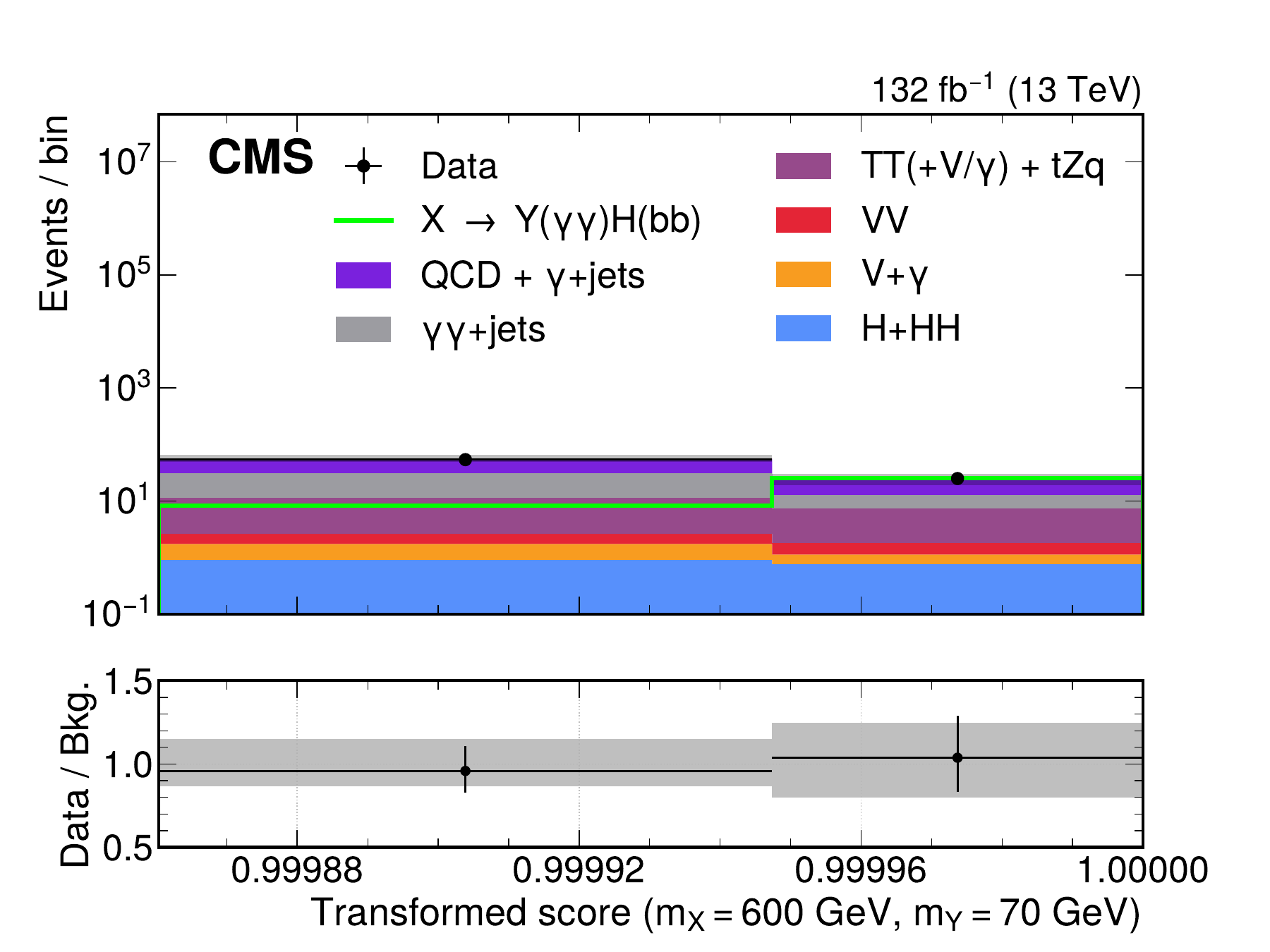}
    \caption{Distributions of the transformed PNN score for the signal hypotheses of $\mX=280\GeV$, $\mY=90\GeV$ (left) and $\mX=600\GeV$, $\mY=70\GeV$ (right) in their corresponding SRs.
    The bin boundaries correspond to the SR boundaries of each mass point, and the evolution of significance with transformed PNN score is shown: for bins with approximately the same background, the signal yield decreases towards the left (less sensitive categories), while the signal yield increases only when the background yield increases as well.
    The distributions are inclusive in the \mgg distribution.
    The gray bands in the lower panels show the statistical uncertainty in the background estimation.}
    \label{fig:pNNScore}
\end{figure}

First, the signal yields are extracted from the simulation in a $\pm 1$ standard deviation window around the signal \mgg peak, and the expected background yields within the same window are extracted from an exponential fit to data in a sideband region, i.e., the \mgg distribution excluding a region $\pm 10\GeV \mY / \mH$ around the signal peak, corresponding to the \mgg resolution.
Next, a constraint is placed on the minimum number of events required in the \mgg sidebands in an SR.
Initially, a minimum number of $N_\text{min}=20$ events is required, ensuring enough statistical power to perform the \mgg sideband fits.
The SR boundaries are defined as follows:

\begin{enumerate}
    \item The first boundary is set at the value of the PNN transformed score which yields  $N = N_\text{min}$ events with scores higher than the boundary.  We then compute for this SR the simplified expected upper limit, defined above, on the $\PX\to\PY(\PGg\PGg)\PH(\PQb\PAQb)$ cross section.
    \item Next, a new SR with the next $N$ highest scoring events is considered, and the simplified expected limit is recomputed, based on the combination of the SRs.
    \item If the addition of a new SR improves the simplified expected limit by at least 5\%, then the new SR is incorporated and Step 2 is repeated for the next SR. Otherwise, this SR is not added and a new SR is formed with twice the number of highest scoring events ($N \to  2N$), and step 2 is repeated.
    The new value of $N$ holds for all subsequent SRs as well.
\end{enumerate}

This procedure continues until the number of events to be added in an SR is larger than the number of preselected events available, \ie,  not having been assigned to a previous SR yet.
A different number of SRs is created at different mass points: mass points with higher \mX-\mY differences tend to have only a few SRs, while mass points with smaller \mX-\mY differences can have up to around 10 SRs.
The last region defined in this procedure has only a small contribution of potential signal and the highest background.
As a result, it is considered as a control region (CR) for the purposes of background estimation, as discussed in Section~\ref{sec:modeling:bkg}.

\section{Signal and background modeling}
\label{sec:modeling}

In this Section we describe the signal and backgrounds PDFs in the extended maximum likelihood fit of the \mgg variable, used to extract the signal cross section, simultaneously across all SRs of a given signal mass hypothesis.
To allow for the different \mY hypotheses, PDFs are defined in a range of $65\GeV < \mgg < 1000\GeV$.
Depending on the value of \mY, data sidebands for nonresonant background fits are constructed around a ``blinding region" defined as:
\begin{equation}\label{eq:blinding_region}
    \mY \pm 10 \GeV \frac{\mY}{\mH}
\end{equation}

\subsection{Signal modeling}
\label{sec:modeling:signal}

\begin{figure} [htb!]
    \centering
    \includegraphics[width=0.49\linewidth]{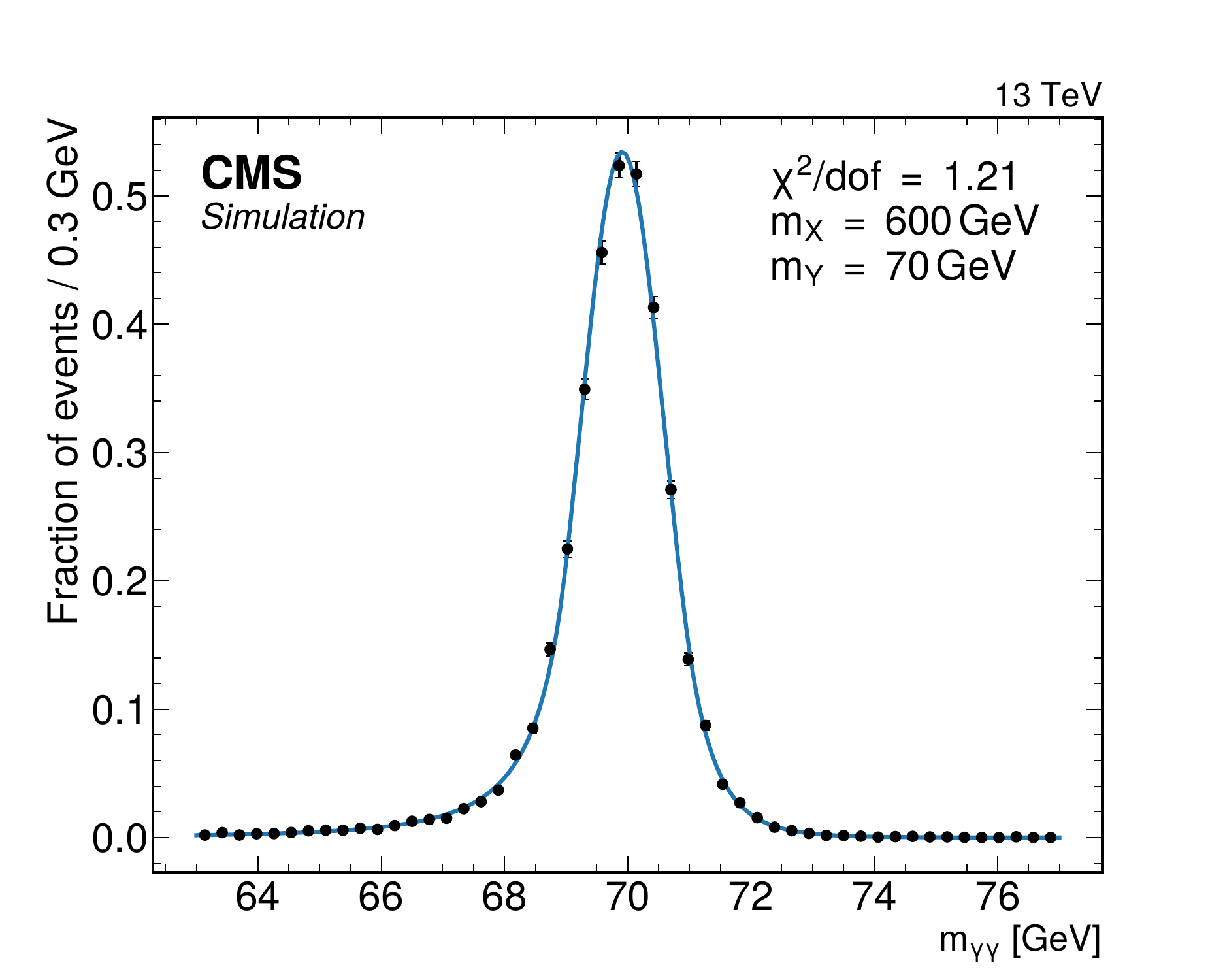}
    \includegraphics[width=0.49\linewidth]{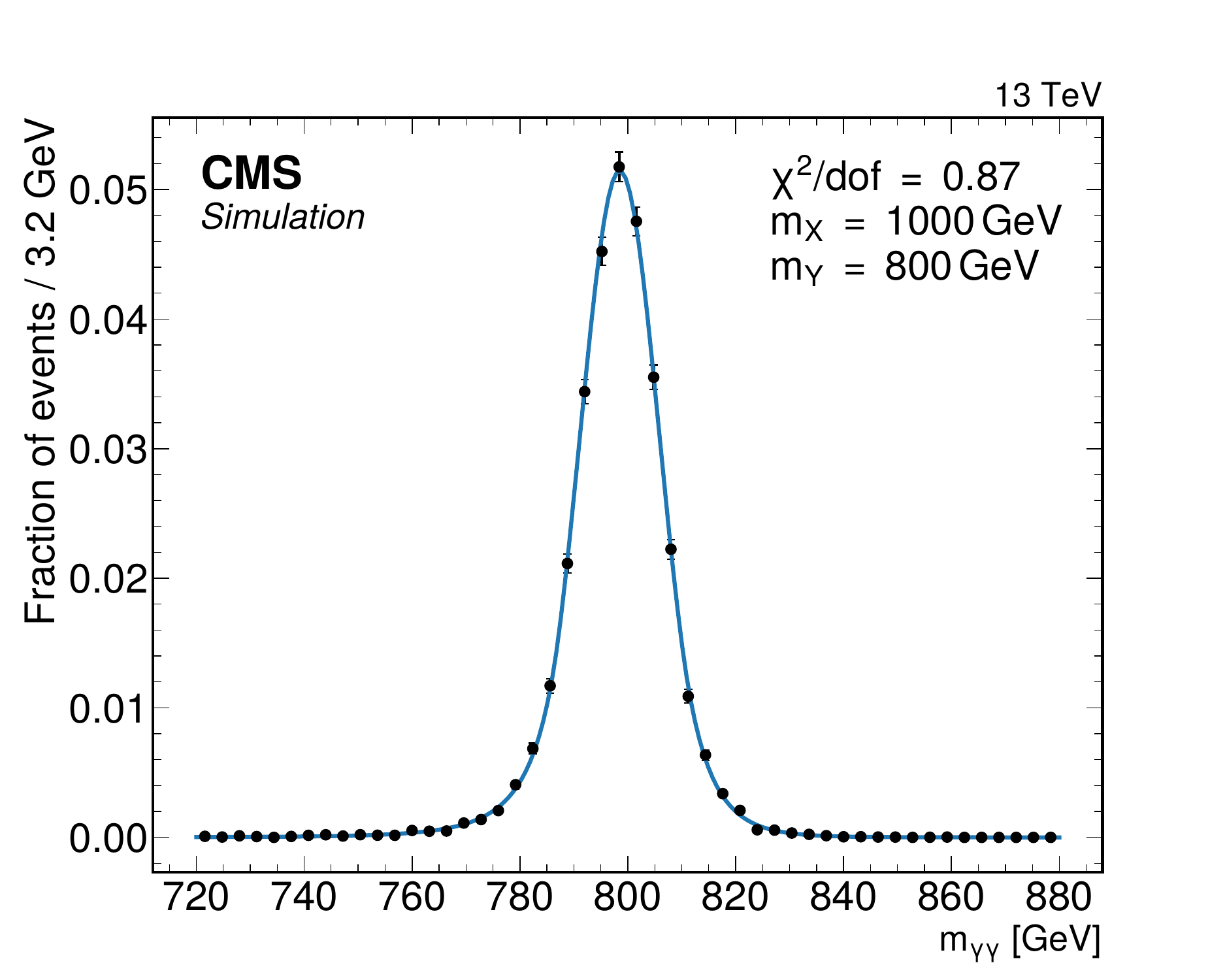}
    \caption{Parametric models of the signal process for $\mX=600\GeV$, $\mY=70\GeV$ (left), and for $\mX=1000\GeV$, $\mY=800\GeV$ (right) in their most sensitive SR. The histograms are normalized to unity. The acronym ``dof" stands for the number of degrees of freedom of the parametric model.}
    \label{fig:smodel}
\end{figure}

The \mgg PDF for signal is taken from simulated signal events after the PNN selection has been applied, independently for each SR and for each year of data taking, since different data-taking periods exhibit up to 10\% difference in the \mgg resolution.
Selected MC signal events consist predominantly of pairs of photons in the barrel part of the detector.
As a result, the resolution for the \mgg peak is driven by the photon resolution in the barrel, and further splitting the events in detector regions does not have an impact on the final result.
For modeling the signal we use double Crystal-Ball (DCB) functions~\cite{Oreglia:1980cs,Gaiser:1982yw}.
Figure~\ref{fig:smodel} shows simulated data fitted with DCB functions for two different mass hypotheses.

\subsection{Background modeling}
\label{sec:modeling:bkg}

Background events are subdivided according to their different \mgg PDFs:
the continuum background arising from nonresonant diphoton events is described by a smoothly falling distribution in \mgg, while other background components peak in \mgg (resonant backgrounds).

The $\PH\to\PGg\PGg$ contribution, from both single \PH and double \PH production, is modeled using the same procedure used for signal events.
Hence, their parametric models are very similar to the ones shown in Fig.~\ref{fig:smodel} for the signal.
The normalization of the different PDFs is constrained to the best available cross section and branching fraction recommendations from the LHC Higgs Working Group~\cite{deFlorian:2016spz}.

A second resonant background contribution consists of DY events when the two electrons from a \PZ boson decay are misidentified as photons, resulting in a peak in the \mgg spectrum close to the \PZ boson mass.
This background is especially important for \mY hypotheses around the \PZ mass, since this contribution can be mistakenly ascribed to a signal, if not properly accounted for.

We use a matrix (``ABCD") method, similar to the one described in Ref.~\cite{Choi2021}, to model this background.
The procedure exploits the electron veto and the PNN output score variables, which are uncorrelated, and is based on three independent CRs in data, as described below:
\begin{itemize}
  \item Region A is defined by inverting both the electron veto and the minimum PNN selection;
  \item Regions B are defined by inverting the electron veto. Multiple regions are defined as each one corresponds to one of the multiple signal regions of the analysis;
  \item Region C is defined by inverting the minimum PNN selection. This is the CR of the analysis, as defined in Section~\ref{sec:selectionCategorization:categorization};
  \item Regions D are the signal regions of the analysis.
\end{itemize}
By inverting the electron veto, regions A and B are enriched in DY events, with \PZ$\to \Pep\Pem$.
Region B constrains the shape of the DY distribution in $\mgg$.
The DY yields are extracted from regions A and C, thereby constraining the electron veto efficiency.

In all the regions, the target $\PZ\rightarrow\Pe^+\Pe^-$ process is isolated by a fitting procedure, with two components: one to model the peaking DY part, described in this paragraph, and one to model the continuum contribution, described in the next paragraph.
The DY \mgg PDFs are assumed to be the same in regions with the same PNN selection (in A and C, and in B and D).
This assumption is justified by the fact that kinematic variables used by the PNN are not correlated with \mgg.
For regions A and C, the PDF is modeled by a Gaussian function, except in cases where a poor goodness-of-fit is found and a DCB function is used instead.
The rates of DY in regions A, B and C are included as free parameters in the maximum likelihood fit.
The rate of the DY background in region D is given by the rate in region B, multiplied by the electron veto efficiency (rate in region C divided by rate in region A).
The ABCD method is applied independently to each SR.
Since the number of SRs depends on the \mX and \mY hypothesis being considered, this implies the definition of a varying number of sub-regions B$_i$ and D$_i$ for each tested hypothesis, while in all cases, regions A and C are uniquely defined.

The procedure for selecting the candidate functions for modeling the \mgg distribution of the continuum background is described below.
The candidate PDFs, extracted separately in each SR of the analysis, are chosen from a set of different analytic functions (Bernstein polynomials, Laurent series, exponentials, and power laws).
For each function type, we determine the minimum and maximum order of parameters by imposing a threshold on the goodness of fit to data and an $\mathcal{F}$-test~\cite{ftest}, respectively.
In the $\mathcal{F}$-test, each function is fit to the \mgg distribution by minimizing the double negative logarithm of the likelihood (\TwoNLL).
To avoid biasing the background model with the eventual signal, fits to the data are performed in \mgg  sidebands.
The size of the \mgg SR depends on \mY following the relation reported in Eq.~\eqref{eq:blinding_region}.
In the fits, the normalization and the function parameters are included as freely floating parameters.
We add a penalty term to the \TwoNLL which is
proportional to the number of function parameters, with the goal to give preference to simple functions over more complex ones.
Finally, we use the discrete profiling procedure~\cite{Dauncey:2014xga} to take into account the systematic uncertainty associated with the choice of the functional form.
In this procedure, a discrete nuisance parameter is associated with the choice of the function.
This nuisance parameter is then profiled in a manner similar to the profiling of the other, continuous, nuisance parameters during the extended maximum likelihood fit of all regions relevant to a specific mass hypothesis.
The potential bias of the nonresonant background estimation arising from a contribution from other signal mass hypotheses outside of the blinding region has been quantified to be of order 1\% and is covered by this uncertainty.

\begin{figure}[hbt!]
	\centering
		\includegraphics[width=0.49\linewidth]{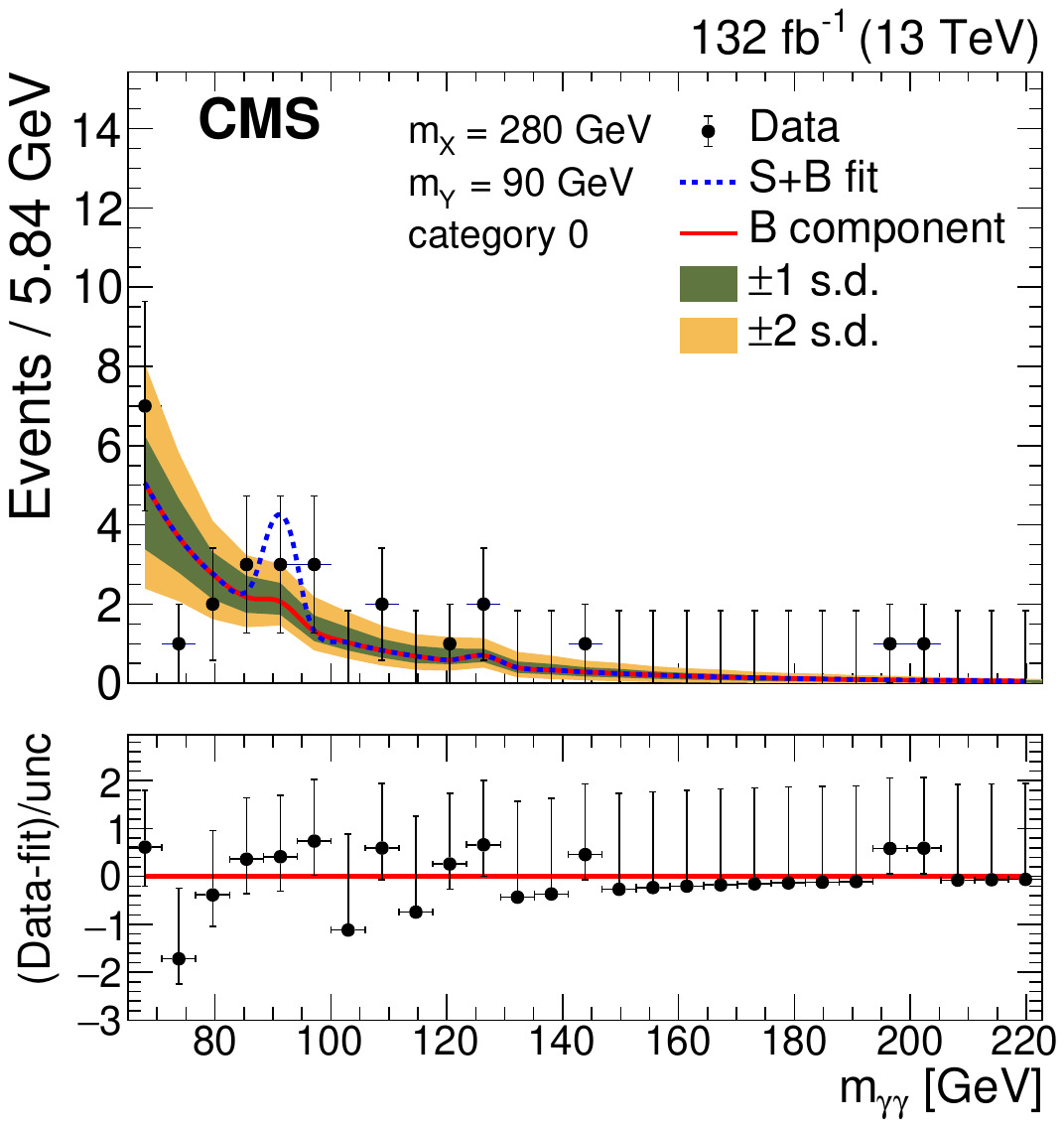}
		\includegraphics[width=0.49\linewidth]{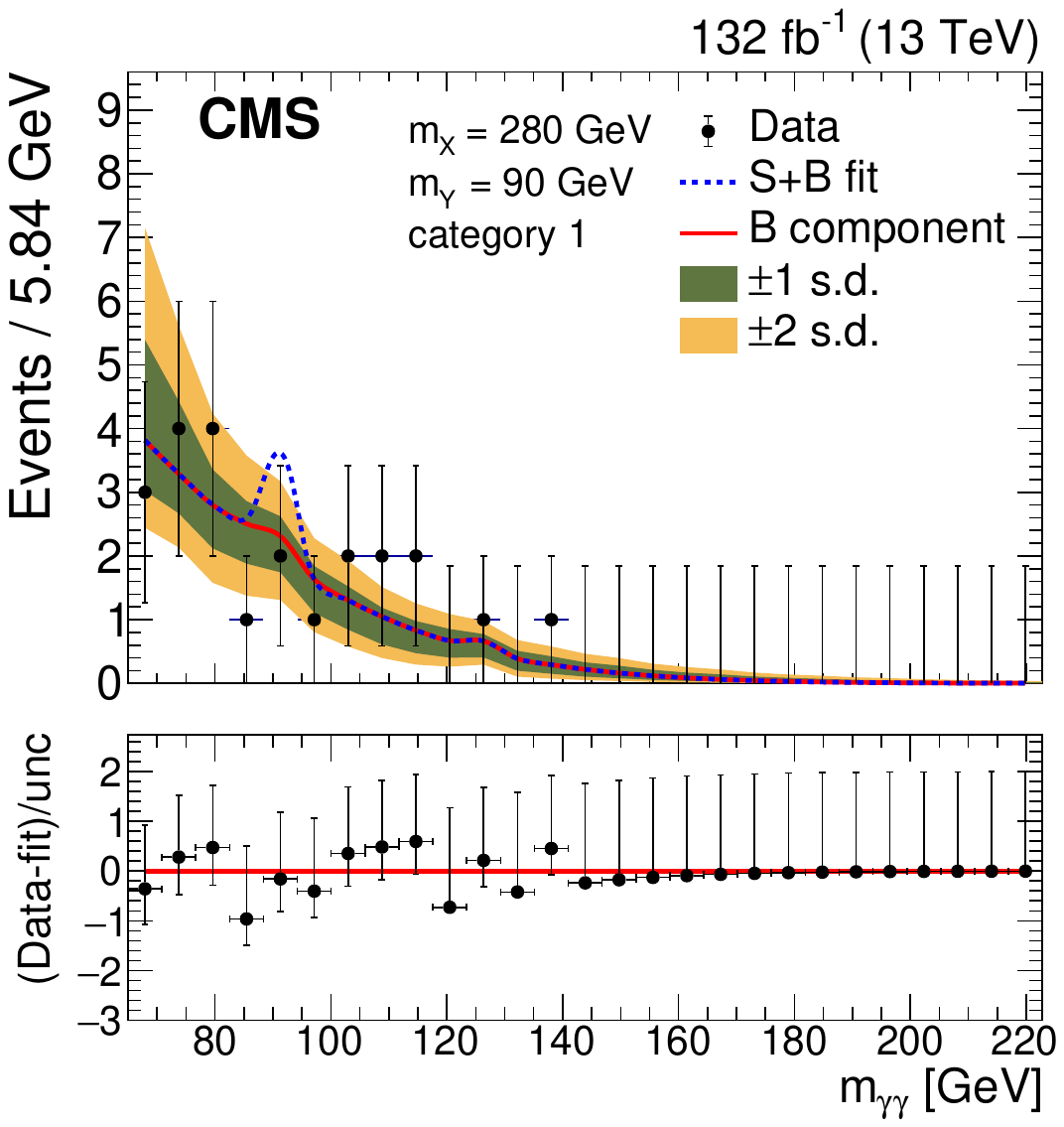}
	\caption{Background-only fit (solid red line) and signal+background fit (dashed blue line) for the mass point hypothesis of $\mX=280\GeV$, $\mY=90\GeV$, shown for the most sensitive SR (left) and the second most sensitive SR (right).
    The points in the lower panel show the difference between the data and the background-only fit, divided by the average uncertainty in the data.
    The red line in the lower panel shows the background-only fit, which is by definition zero, and it is added as a visual aid.
    From left to right, the first and second most sensitive SRs are shown.
    The choice of the background functional form is determined by the maximum likelihood fit.}
	\label{fig:bmodel_nonres}
\end{figure}

As an illustration of the different components of our background estimation (continuum background from nonresonant processes and the \PZ and \PH resonances), in Fig.~\ref{fig:bmodel_nonres} we show the background-only fit and signal+background fit for $\mX=280\GeV$, $\mY=90\GeV$ hypotheses in the two most sensitive SRs for this mass point.

\section{Systematic uncertainties}
\label{sec:systematics}

The sensitivity of the search is driven by the integrated luminosity of the data set and the small number of events in the SRs.  We do, however, account for a comprehensive set of systematic uncertainties, as described in this section.

The systematic uncertainties affecting the simulated samples used in the analysis (signal and Higgs production) are incorporated into the final maximum likelihood fit as nuisance parameters (NPs), and are described in detail below.
The majority of these systematic uncertainties affect the overall expected event yields for both the signal and the single Higgs production. These uncertainties are modeled with log-normal probability distribution functions.
For these processes there are also uncertainties in the measurements of the photon energies.  These uncertainties impact the shape of the corresponding \mgg distributions, and they are incorporated as NPs on the mean and the width of the \mgg PDFs.

As discussed in Section~\ref{sec:modeling:bkg}, the systematic uncertainty associated with the nonresonant background modeling is taken into account using the discrete profiling method.
Finally, we use uniform priors for the transfer factors that determine the normalization of the resonant DY background.

\subsection{Theoretical uncertainties}
\label{sec:systematics:theory}

A number of theoretical uncertainties affect the normalization of the single and pair SM Higgs boson production processes:

\begin{itemize}
    \item Theoretical uncertainties in the $\PH\PH$ cross section:
    \begin{itemize}
        \item The parton distribution functions, and the value of the strong force coupling constant ($\alpha_s$), are estimated within the Born-improved approximation based on the PDF4LHC15nnlomc set~\cite{pdf4lhc}. They are found to be ${\pm}3.0\%$.
        \item Uncertainties arising from the choice of the renormalization scheme and the scale of the top quark mass, and their combination with renormalization, $\mu_\text{R}$ and factorization, $\mu_\text{F}$, scale variations are based on Ref.~\cite{Baglio_2021}. They are found to be ${}^{+6\%}_{-23\%}$.
    \end{itemize}
    \item Theoretical uncertainties in the single \PH boson production cross section:
    \begin{itemize}
        \item QCD $\mu_\text{R}$ and $\mu_\text{F}$ scale variation values are taken following the recommendations of Ref.~\cite{deFlorian:2016spz}, and typically result in an uncertainty of the order of 5--10\%.
        \item The $\alpha_s$ variations are handled following the recommendations of Ref.~\cite{pdf4lhc} and result in an up to 2\% uncertainty in the cross section.
        \item The uncertainty associated with the parton distribution functions is computed following the PDF4LHC prescription~\cite{pdf4lhc,higgs_handbook}, and is typically of the order of 1\%.
    \end{itemize}
\end{itemize}

Finally, we consider the uncertainties in the branching fraction $\mathcal{B}(\PH\to\PGg\PGg)$ for both the single and double SM \PH processes.
These are taken from Ref.~\cite{deFlorian:2016spz} and are estimated to be around 2\%.

\subsection{Experimental uncertainties}
\label{sec:systematics:experiment}

The experimental uncertainties that affect the normalization of the simulated  signal and Higgs production samples are:

\begin{itemize}
    \item Integrated luminosity:
    The integrated luminosities for the 2016, 2017, and 2018 data-taking years have 1.2--2.5\% individual uncertainties~\cite{CMS-LUM-17-003,CMS-PAS-LUM-17-004,CMS-PAS-LUM-18-002}, while the overall uncertainty for the 2016--2018 period is 1.6\%.
    Since the signal and resonant \PH background fits are performed separately for the three data-taking periods, we use the year-dependent uncertainties and their correlations.
    \item Pileup:
    The uncertainties derived from the reweighting procedure to match the simulated pileup distribution to the one in the data, and originating from the variation of the minimum bias cross section~\cite{PU}, lead to an impact on the final results of less than 1\%.
    \item Trigger:
    Trigger efficiencies are measured with $\PZ\to \Pep\Pem$ data using the ``tag-and-probe" method~\cite{CMS:2011aa}.
    Their uncertainties are propagated throughout the analysis, including those associated with a small loss of efficiency due to a gradual shift in the ECAL trigger timing during the 2016 and 2017 data-taking periods.
    The size of the trigger-related uncertainties combined is less than 1\%.
    \item Photon and diphoton identification: These uncertainties affect the final result at the 2\% level.
    \item Jet energy correction (JEC) and jet energy resolution (JER):
    Eleven separate sources of JEC uncertainties, as well as the uncertainty arising from the JERs, are propagated separately to the properties of jets in simulation~\cite{CMS:2016lmd}.
 These result in 12 systematic variations per year, which are propagated into the final result.
    The maximum impact of these uncertainties ranges between 0.7\% and 3.0\%, depending on the signal mass point.
    \item \PQb tagging:
    The \textsc{DeepJet} output score is reshaped to correct for the differences between data and simulation~\cite{BTV-16-002}.
    The corresponding uncertainty sources involved in this procedure are included in the analysis, taking into account the appropriate correlation scheme.
    Nine separate sources of uncertainty are considered. We find the impact of these uncertainties to be less than 1\% for each uncertainty source.
\end{itemize}

The experimental uncertainties, affecting the shape of the \mgg distribution, are evaluated with the ``tag-and-probe" method with $\PZ\to \Pep\Pem$ data~\cite{CMS:2011aa}. These uncertainties are:

\begin{itemize}
    \item Photon energy scale and resolution:
    Corrections are applied to simulation to match its photon energy scale and resolution to that of data.
    The size of these uncertainties combined is less than 1\%.
    \item Tracker and ECAL material uncertainties:
    Electrons showers tend to develop earlier than those of photons in the material of the tracker.
    Light collection efficiency differs for electrons and photons along the length of the ECAL crystals, and this leads to different ECAL responses.
    The uncertainty in the responses arising from these two effects is considered in the analysis, leading to an impact on the final results of approximately 1\%.
\end{itemize}

As already mentioned at the beginning of the section, the sensitivity is dominated by the statistical uncertainties.
The statistical uncertainty ranges from 30--50\%, depending on the mass point, and the inclusion of all the systematic uncertainties increases the total uncertainty by up to 2\% in absolute terms.

\section{Results}
\label{sec:results}

The data are found to be compatible with the background-only expectation for most mass hypotheses probed in this search.
We therefore extract upper limits on the process $\PX\rightarrow\PY(\PGg\PGg)\PH(\PQb\PAQb)$.
The most significant deviation from the background-only hypothesis is discussed below.

\begin{figure}[htbp]
    \centering
    \includegraphics[width=0.78\textwidth]{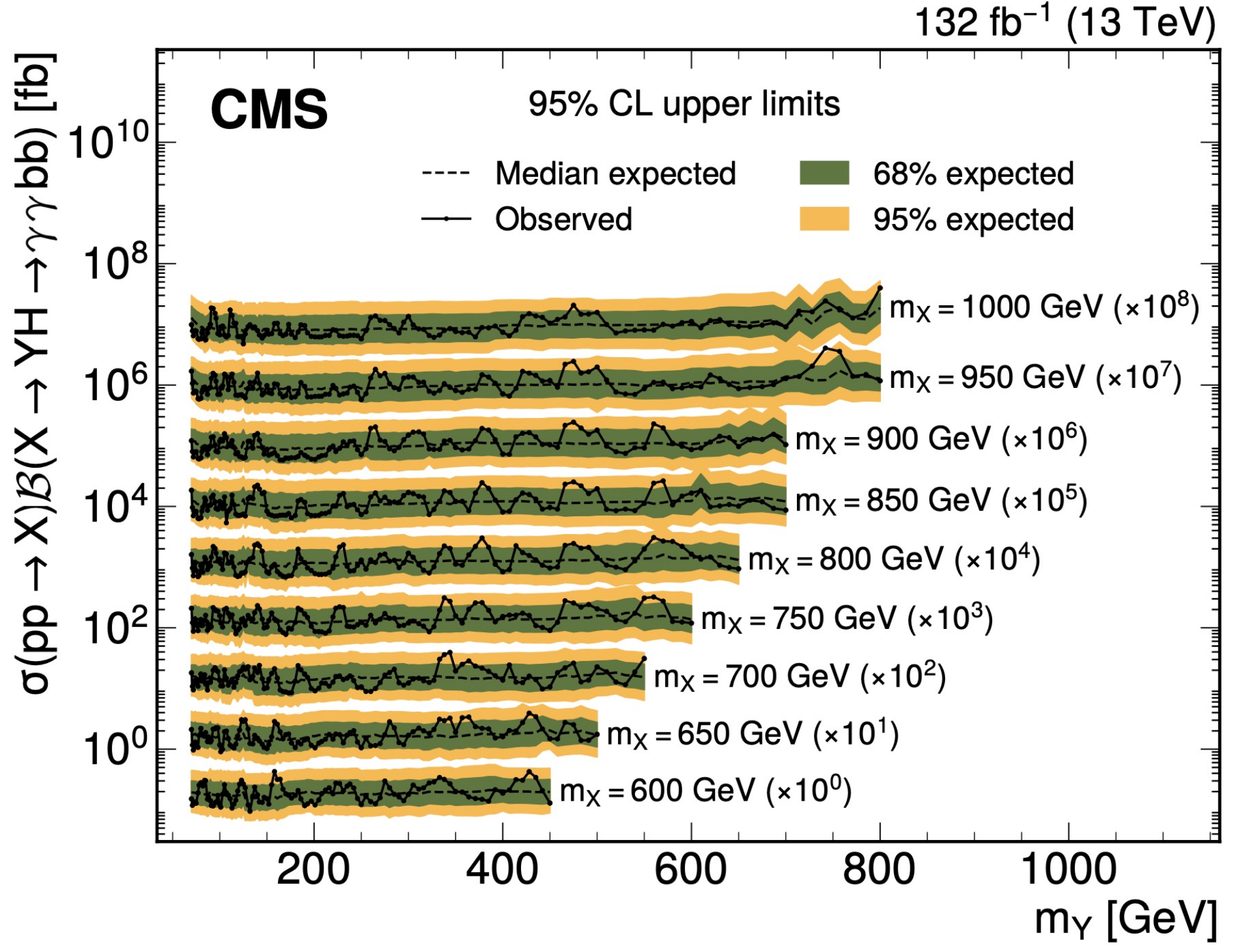}
    \includegraphics[width=0.78\textwidth]{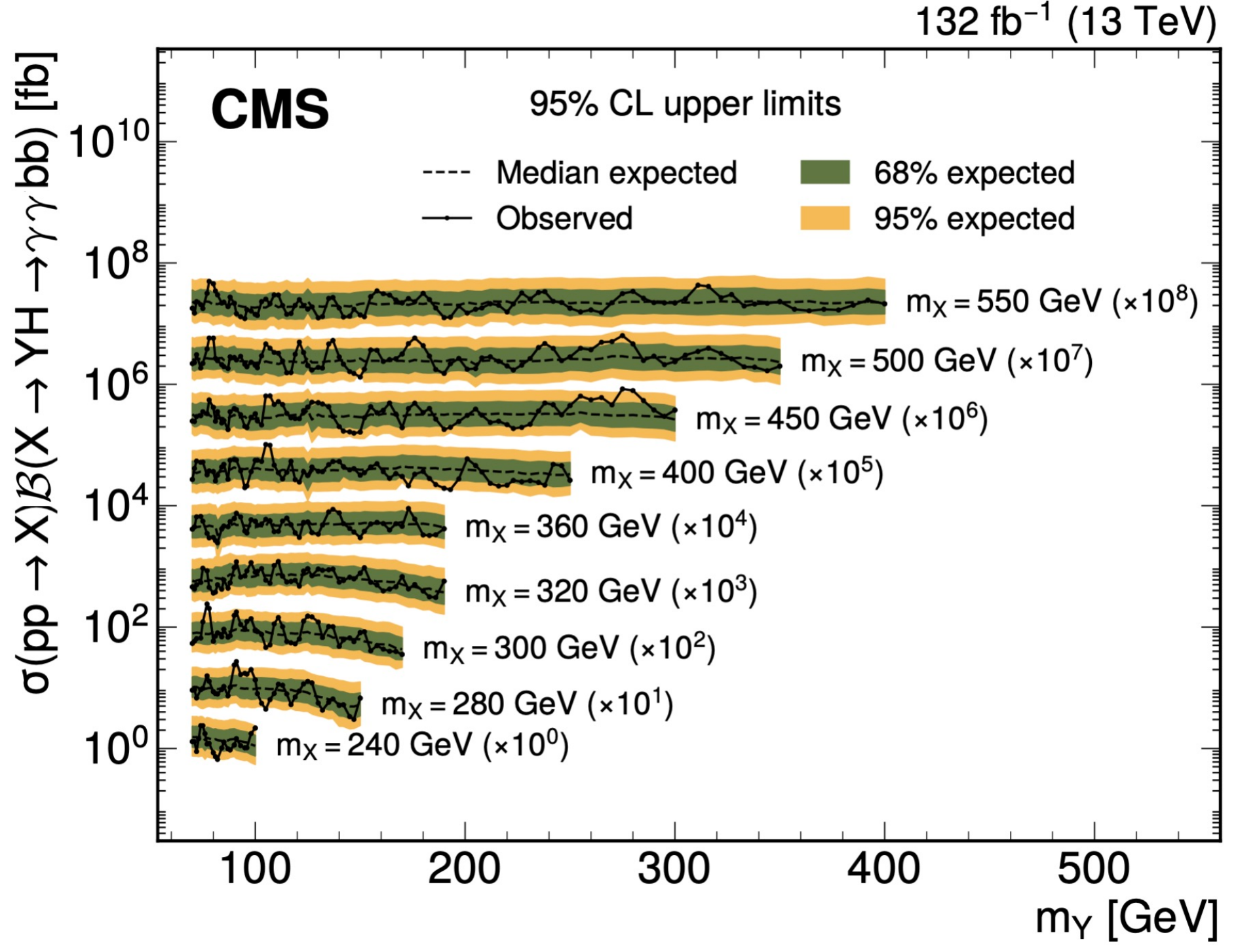}
    \caption{Observed (expected) upper limits on $\sigma\mathcal{B}$ for the $\PX\to\PY(\PGg\PGg)\PH(\PQb\PAQb)$ signal with the different mass hypotheses are shown with solid (dashed) lines, for mass points with \mX ranging from 600 to 1000\GeV (upper) and from 240 to 550\GeV (lower).
    The green and the yellow bands define the $\pm$68\% and $\pm$95\% uncertainty bands, respectively.
    For visualization purposes, the upper limits for mass points with different \mX have been multiplied with a constant factor quoted on the right of each band.}
    \label{fig:stackedLimitPlots}
\end{figure}

\begin{figure}[htbp]
    \centering
    \includegraphics[width=0.49\textwidth]{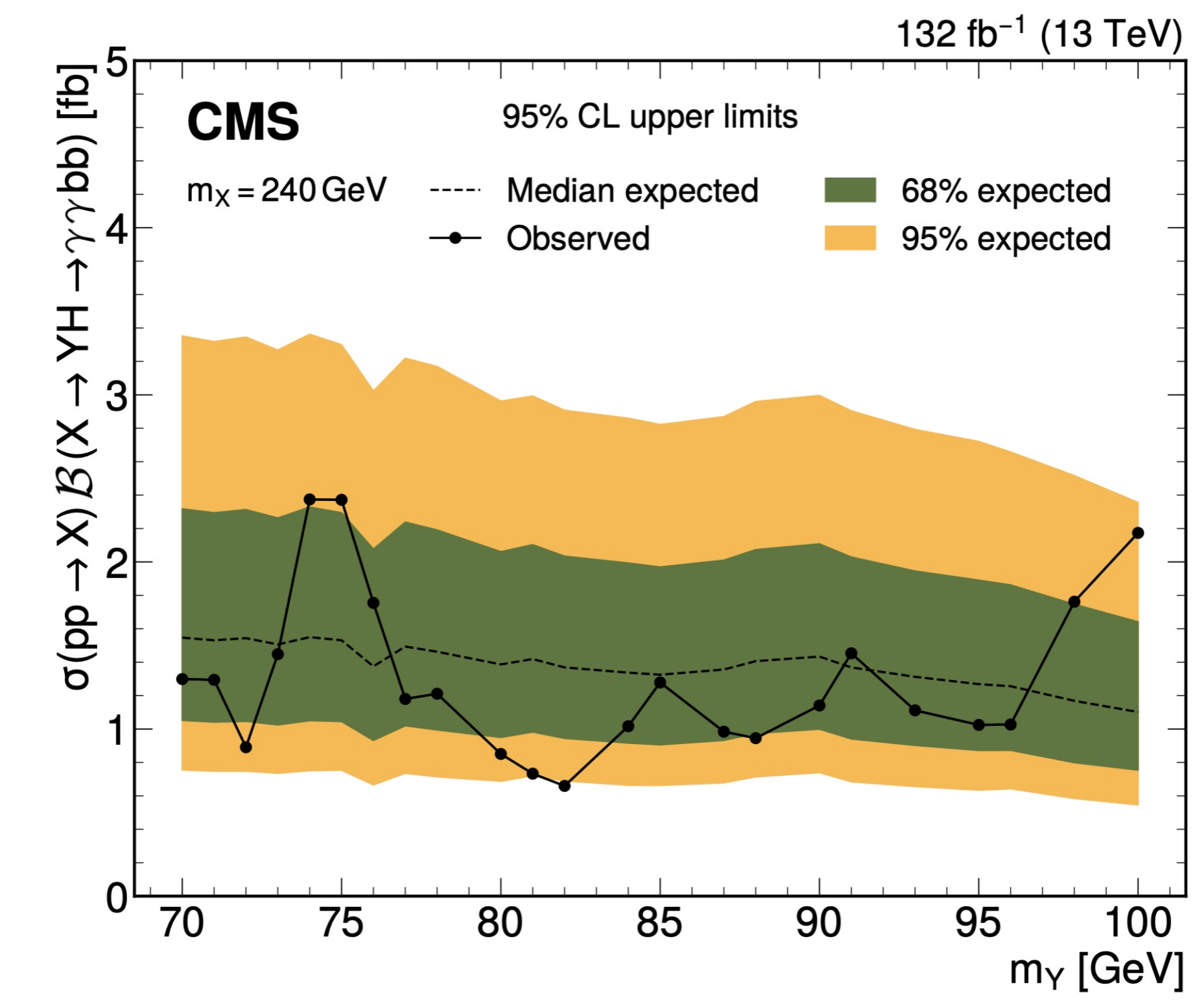}
    \includegraphics[width=0.49\textwidth]{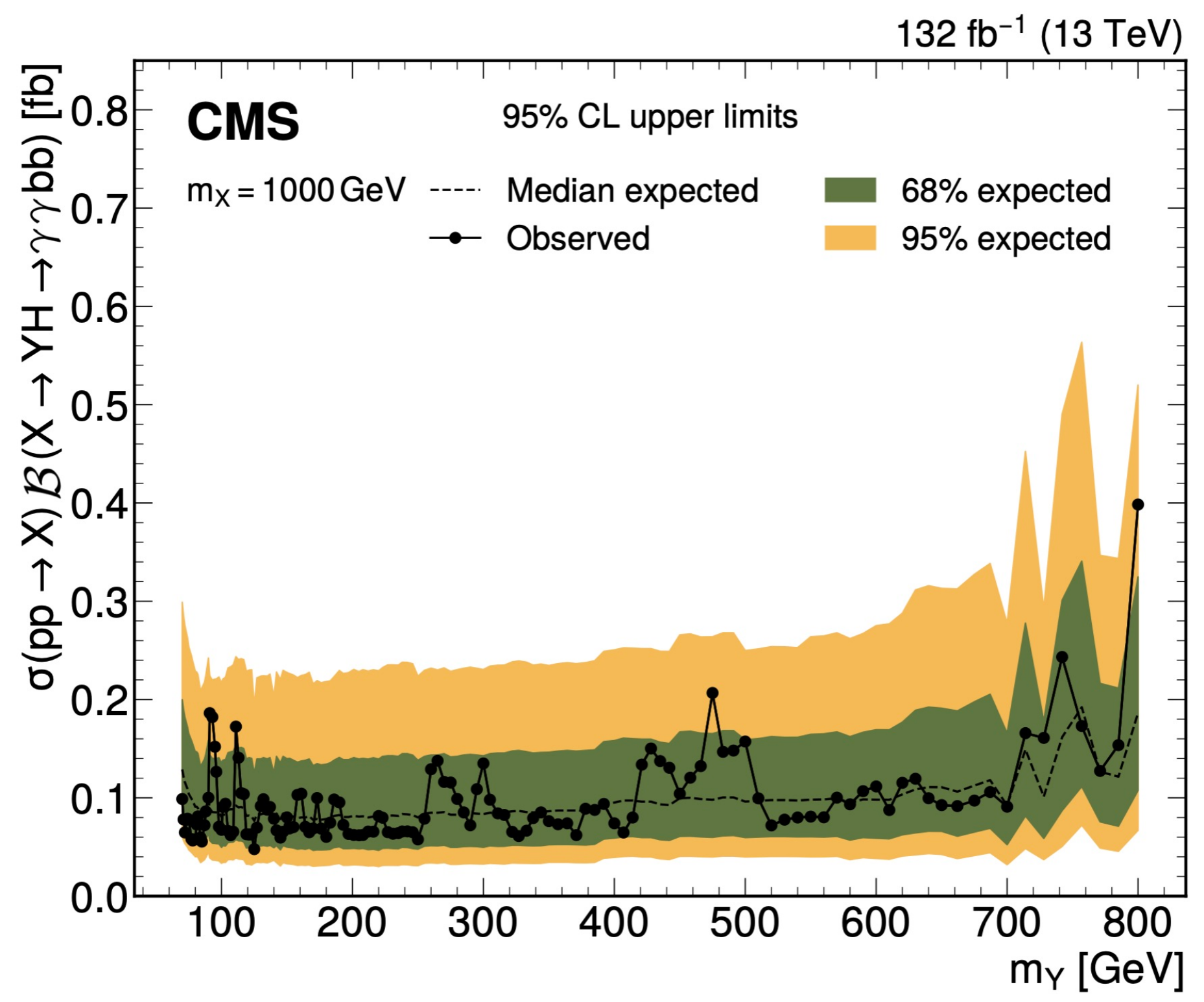}
    \caption{Observed (expected) upper limits on $\sigma\mathcal{B}$ for the $\PX\to\PY(\PGg\PGg)\PH(\PQb\PAQb)$ signal with the different \mY hypotheses are shown with solid (dashed) lines, shown for the lowest \mX = 240\GeV (left) and for the highest \mX = 1000\GeV (right). The inner (green) band and the outer (yellow) band indicate the regions containing 68\% and 95\%, respectively, of the distribution of limits expected under the background-only hypothesis.}
    \label{fig:expectedLimitPlot_mX240_mX1000}
\end{figure}

The results quoted in this section have been determined using the CMS statistical analysis tool \textsc{Combine}~\cite{CMS:2024onh}, which is based on the \textsc{RooFit}~\cite{Verkerke:2003ir} and \textsc{RooStats}~\cite{Moneta:2010pm} frameworks.
Figure~\ref{fig:stackedLimitPlots} shows the summary of expected and observed upper limits on the product, $\sigma\mathcal{B}$, of the signal cross section and the corresponding $\mathcal{B}$ to the final state particles for the different \mX and \mY values considered in the analysis.
Figure~\ref{fig:expectedLimitPlot_mX240_mX1000} illustrates the same expected and observed upper limits for the lowest and the highest sets of \mX mass points.
While extracting results, both ``nominal" mass points, i.e., the mass points on whose signal simulation the PNN has been trained, and ``interpolated" mass points, i.e., mass points for which there is no simulation available, have been used.
All of the results have been derived using the signal and background modeling methods described in detail in Section~\ref{sec:modeling}.
Systematic uncertainties are applied according to their description in Section~\ref{sec:systematics}.

To achieve a sensitivity on the interpolated mass points comparable to that of the nominal mass points, we have leveraged the interpolating capabilities of the PNN.
A dedicated PNN output is produced for each mass point that is used for the event categorization and the extraction of the results for the different signal hypotheses.
Selected according to the dedicated PNN outputs, the data and the background simulation are used to extract the background models for the interpolated mass points, as described in Section~\ref{sec:modeling:bkg}, while, in the absence of signal simulation for those mass points, the signal model parameters are interpolated between nominal mass points using linear splines.
The interpolated mass points have been inserted at intervals corresponding to approximately twice the resolution at the given \mgg range, i.e., ranging from 1\GeV up to 8\GeV.

For a fixed \mX value, the limits tend to be weaker for low \mY values ($\mY < 200$\GeV) owing to the larger  backgrounds in that phase space region, and the presence of SM resonances.
At higher \mY values, the limits increase for mass points with lower $\mX-\mY$ difference.
This is attributed to the lower efficiency in distinguishing signal from background events in this phase space region.
In contrast, for mass points with higher $\mX-\mY$ difference, the observed limits are generally lower, reflecting the improved signal-to-background ratio.
This behavior is a direct consequence of the different kinematic regions explored.

This is the first time that the $\PX\to\PY(\PGg\PGg)\PH(\PQb\PAQb)$ process is probed, i.e., with the resonance \PY, which is of unknown mass, decaying to a photon pair.
Given that the diphoton mass peak is usually exploited to tag the signal in similar final states, a moving signal peak poses additional experimental challenges.
Despite these, the results of this analysis in terms of upper limits are comparable with those reported in the search for $\PX\to\PY(\PQb\PAQb)\PH(\PGg\PGg)$~\cite{XToYbbHgg}.

In a recent $\PX\to\PY(\PQb\PAQb)\PH(\PGg\PGg)$ search, there was a local (global) excess of significance of 3.8\,(2.6) standard deviations at $\mX = 650\GeV$ and $\mY = 90\GeV$. However, in the current analysis, the observed limit at $\mX = 650\GeV$ and $\mY = 90\GeV$ is in agreement with the SM, falling within 1 standard deviation of the median expected value.
Under the assumption that the branching fractions of the \PY to photons and \PQb quarks are the same as the ones of the Higgs boson, as in the TRSM, the observed upper limit on $\sigma\mathcal{B}$ is 0.13\unit{fb}, which is compatible within two standard deviations with the best fit value of the excess of the $\PX\to\PY(\PQb\PAQb)\PH(\PGg\PGg)$ analysis at the same mass point, which was measured to be $0.35\pm^{0.17}_{0.13}\unit{fb}$.
Thus the results of this analysis do not confirm the excess.

In general, the data are found to be compatible with the SM expectation.
A local significance of 3.33 standard deviations has been observed for the most significant deviation from the background-only hypothesis, at a mass point where the \PX and \PY particles have a mass of 300\GeV and 77\GeV, respectively.
Taking into account the large number (1699) of interpolated mass points probed in estimating the look-elsewhere effect~\cite{lookElsewhereEffect}, which refers to the probability of observing a fluctuation of equal or greater significance anywhere in the scanned (\mX,\mY) mass plane, the global significance decreases to 0.65 standard deviations.

\section{Summary}
\label{sec:summary}

A search has been presented for the production of a new scalar resonance, \PX, decaying to a standard model Higgs boson and a new scalar resonance, \PY, with the final state including a $\PQb\PAQb$ pair from the Higgs boson decay and a pair of photons from the \PY particle decay.
This search is the first targeting this final state combination.
The analysis probes the mass range {240--1000\GeV} for the resonance X and {70--800\GeV} for the particle \PY, and uses proton-proton collision data collected by the CMS experiment at $\sqrt{s} = 13\TeV$, corresponding to an integrated luminosity of 132\fbinv.

In general, the data are found to be compatible with the background-only expectation.
As a result, upper limits on the product of the production cross section of \PX and the branching fraction to the $\PQb\PAQb\PGg\PGg$ final state  are derived at 95\% confidence level as functions of the masses of the \PX and the \PY particles.
The observed\,(expected) upper limits are found to be between 0.05--2.69\,(0.08--1.94)\unit{fb}, depending on the assumed specific signal masses.
A local\,(global) significance of 3.33\,(0.65) standard deviations is observed for the most significant deviation from the background-only hypothesis, corresponding to $\mX=300\GeV$ and $\mY=77\GeV$.

\begin{acknowledgments}
We congratulate our colleagues in the CERN accelerator departments for the excellent performance of the LHC and thank the technical and administrative staffs at CERN and at other CMS institutes for their contributions to the success of the CMS effort. In addition, we gratefully acknowledge the computing centres and personnel of the Worldwide LHC Computing Grid and other centres for delivering so effectively the computing infrastructure essential to our analyses. Finally, we acknowledge the enduring support for the construction and operation of the LHC, the CMS detector, and the supporting computing infrastructure provided by the following funding agencies: SC (Armenia), BMBWF and FWF (Austria); FNRS and FWO (Belgium); CNPq, CAPES, FAPERJ, FAPERGS, and FAPESP (Brazil); MES and BNSF (Bulgaria); CERN; CAS, MoST, and NSFC (China); MINCIENCIAS (Colombia); MSES and CSF (Croatia); RIF (Cyprus); SENESCYT (Ecuador); ERC PRG, TARISTU24-TK10 and MoER TK202 (Estonia); Academy of Finland, MEC, and HIP (Finland); CEA and CNRS/IN2P3 (France); SRNSF (Georgia); BMBF, DFG, and HGF (Germany); GSRI (Greece); NKFIH (Hungary); DAE and DST (India); IPM (Iran); SFI (Ireland); INFN (Italy); MSIT and NRF (Republic of Korea); MES (Latvia); LMTLT (Lithuania); MOE and UM (Malaysia); BUAP, CINVESTAV, CONACYT, LNS, SEP, and UASLP-FAI (Mexico); MOS (Montenegro); MBIE (New Zealand); PAEC (Pakistan); MES, NSC, and NAWA (Poland); FCT (Portugal); MESTD (Serbia); MICIU/AEI and PCTI (Spain); MOSTR (Sri Lanka); Swiss Funding Agencies (Switzerland); MST (Taipei); MHESI and NSTDA (Thailand); TUBITAK and TENMAK (T\"{u}rkiye); NASU (Ukraine); STFC (United Kingdom); DOE and NSF (USA).

\hyphenation{Rachada-pisek} Individuals have received support from the Marie-Curie programme and the European Research Council and Horizon 2020 Grant, contract Nos.\ 675440, 724704, 752730, 758316, 765710, 824093, 101115353, 101002207, 101001205, and COST Action CA16108 (European Union); the Leventis Foundation; the Alfred P.\ Sloan Foundation; the Alexander von Humboldt Foundation; the Science Committee, project no. 22rl-037 (Armenia); the Fonds pour la Formation \`a la Recherche dans l'Industrie et dans l'Agriculture (FRIA-Belgium); the Beijing Municipal Science \& Technology Commission, No. Z191100007219010, the Fundamental Research Funds for the Central Universities, the Ministry of Science and Technology of China under Grant No. 2023YFA1605804, and the Natural Science Foundation of China under Grant No. 12061141002 (China); the Ministry of Education, Youth and Sports (MEYS) of the Czech Republic; the Shota Rustaveli National Science Foundation, grant FR-22-985 (Georgia); the Deutsche Forschungsgemeinschaft (DFG), among others, under Germany's Excellence Strategy -- EXC 2121 ``Quantum Universe" -- 390833306, and under project number 400140256 - GRK2497; the Hellenic Foundation for Research and Innovation (HFRI), Project Number 2288 (Greece); the Hungarian Academy of Sciences, the New National Excellence Program - \'UNKP, the NKFIH research grants K 131991, K 133046, K 138136, K 143460, K 143477, K 146913, K 146914, K 147048, 2020-2.2.1-ED-2021-00181, TKP2021-NKTA-64, and 2021-4.1.2-NEMZ\_KI-2024-00036 (Hungary); the Council of Science and Industrial Research, India; ICSC -- National Research Centre for High Performance Computing, Big Data and Quantum Computing, FAIR -- Future Artificial Intelligence Research, and CUP I53D23001070006 (Mission 4 Component 1), funded by the NextGenerationEU program (Italy); the Latvian Council of Science; the Ministry of Education and Science, project no. 2022/WK/14, and the National Science Center, contracts Opus 2021/41/B/ST2/01369, 2021/43/B/ST2/01552, 2023/49/B/ST2/03273, and the NAWA contract BPN/PPO/2021/1/00011 (Poland); the Funda\c{c}\~ao para a Ci\^encia e a Tecnologia, grant CEECIND/01334/2018 (Portugal); the National Priorities Research Program by Qatar National Research Fund; MICIU/AEI/10.13039/501100011033, ERDF/EU, "European Union NextGenerationEU/PRTR", and Programa Severo Ochoa del Principado de Asturias (Spain); the Chulalongkorn Academic into Its 2nd Century Project Advancement Project, and the National Science, Research and Innovation Fund via the Program Management Unit for Human Resources \& Institutional Development, Research and Innovation, grant B39G680009 (Thailand); the Kavli Foundation; the Nvidia Corporation; the SuperMicro Corporation; the Welch Foundation, contract C-1845; and the Weston Havens Foundation (USA).
\end{acknowledgments}

\bibliography{auto_generated}
\cleardoublepage \appendix\section{The CMS Collaboration \label{app:collab}}\begin{sloppypar}\hyphenpenalty=5000\widowpenalty=500\clubpenalty=5000\input{B2G-24-001-public-authorlist.tex}\end{sloppypar}
\end{document}

%% file: B2G-24-001-public-authorlist.tex
\cmsinstitute{Yerevan Physics Institute, Yerevan, Armenia}
{\tolerance=6000
A.~Hayrapetyan, V.~Makarenko\cmsorcid{0000-0002-8406-8605}, A.~Tumasyan\cmsAuthorMark{1}\cmsorcid{0009-0000-0684-6742}
\par}
\cmsinstitute{Institut f\"{u}r Hochenergiephysik, Vienna, Austria}
{\tolerance=6000
W.~Adam\cmsorcid{0000-0001-9099-4341}, J.W.~Andrejkovic, L.~Benato\cmsorcid{0000-0001-5135-7489}, T.~Bergauer\cmsorcid{0000-0002-5786-0293}, M.~Dragicevic\cmsorcid{0000-0003-1967-6783}, C.~Giordano, P.S.~Hussain\cmsorcid{0000-0002-4825-5278}, M.~Jeitler\cmsAuthorMark{2}\cmsorcid{0000-0002-5141-9560}, N.~Krammer\cmsorcid{0000-0002-0548-0985}, A.~Li\cmsorcid{0000-0002-4547-116X}, D.~Liko\cmsorcid{0000-0002-3380-473X}, M.~Matthewman, I.~Mikulec\cmsorcid{0000-0003-0385-2746}, J.~Schieck\cmsAuthorMark{2}\cmsorcid{0000-0002-1058-8093}, R.~Sch\"{o}fbeck\cmsAuthorMark{2}\cmsorcid{0000-0002-2332-8784}, D.~Schwarz\cmsorcid{0000-0002-3821-7331}, M.~Shooshtari, M.~Sonawane\cmsorcid{0000-0003-0510-7010}, W.~Waltenberger\cmsorcid{0000-0002-6215-7228}, C.-E.~Wulz\cmsAuthorMark{2}\cmsorcid{0000-0001-9226-5812}
% ...
\par}
\vskip\cmsinstskip
\dag:~Deceased\\
$^{1}$Also at Yerevan State University, Yerevan, Armenia\\
$^{2}$Also at TU Wien, Vienna, Austria\\
% ...
